\documentclass[twocolumn, tighten, times, twocolappendix]{aastex631}
\usepackage{amsmath,amstext}
\usepackage{xcolor,soul}
\makeatletter %% <- make @ usable in command sequences
\newcount\SOUL@minus
\makeatother
\usepackage[figure,figure*]{hypcap}
\usepackage{multirow}
\usepackage[T1]{fontenc} 
\usepackage[utf8]{inputenc} 
\usepackage{float}
\usepackage{tikz}
\newcommand{\ba}{\begin{eqnarray}}
\newcommand{\ea}{\end{eqnarray}}
\renewcommand*{\l}{\ensuremath{\left}}
\renewcommand*{\r}{\ensuremath{\right}}
\newcommand*{\f}{\ensuremath{\frac}}
\newcommand*{\p}{\ensuremath{\partial}}
\newcommand*{\oB}{\ensuremath{_\text{out}}}
\newcommand*{\iB}{\ensuremath{_\text{in}}}
\newcommand*{\bin}{\ensuremath{_\text{bin}}}
\newcommand*{\BH}{\ensuremath{_\text{BH}}}
\newcommand*{\cl}{\ensuremath{_\text{cl}}}
\newcommand*{\GR}{\ensuremath{_\text{GR}}}

\renewcommand*{\L}{\ensuremath{\mathcal{L}}}
\newcommand*{\G}{\ensuremath{\mathcal{G}}}
\renewcommand*{\H}{\ensuremath{\mathcal{H}}}
\newcommand{\mthspc}
\shorttitle{Binary evolution perturbed by flybys and tidal fields}
\shortauthors{M. Winter-Granic, C. Petrovich, V. Peña-Donaire \& C. Hamilton}

\begin{document}

\title{Binary mergers in the centers of galaxies:  synergy between stellar flybys and tidal fields}

\correspondingauthor{Mila Winter-Granic}
\email{mila.winter@uc.cl}

\author[0009-0005-8908-4312]{Mila Winter-Granic}
\affiliation{Instituto de Astrofísica, Pontificia Universidad Católica de Chile, Av. Vicuña Mackenna 4860, 782-0436 Macul, Santiago, Chile}

\author[0000-0003-0412-9314]{Cristobal Petrovich}
\affiliation{Instituto de Astrofísica, Pontificia Universidad Católica de Chile, Av. Vicuña Mackenna 4860, 782-0436 Macul, Santiago, Chile}
\affiliation{Millennium Institute of Astrophysics MAS, Nuncio Monseñor Sótero Sanz 100, Of. 104, 750-0000 Providencia, Santiago, Chile}
\affiliation{Department of Astronomy, Indiana University, Bloomington, IN 47405, USA}
\author[0000-0003-2341-3134]{Valentín Peña-Donaire}
\affiliation{Instituto de Astrofísica, Pontificia Universidad Católica de Chile, Av. Vicuña Mackenna 4860, 782-0436 Macul, Santiago, Chile}

\author[0000-0002-5861-5687]{Chris Hamilton}
\affiliation{Institute for Advanced Study, Einstein Drive, Princeton, NJ 08540, USA}

%\affiliation{Centro de Astrofísica y Tecnologías Afines CATA, Camino El Observatorio 1515, 759-1245 Las Condes, Santiago, Chile}
%Bibliographical info

\begin{abstract}
\noindent
Galactic centers are very dynamically active environments, often harbouring a nuclear star cluster and supermassive black hole at their cores. Binaries in these environments are subject to strong tidal fields that can efficiently torque its orbit, exciting near unity eccentricities that ultimately lead to their merger. In turn, frequent close interactions with passing stars impulsively perturb the orbit of the binary, generally softening their orbit until their evaporation, potentially hindering the role of tides to drive these mergers. In this work, we study the evolution of compact object binaries in the galactic center and their merger rates, focusing for the first time on the combined effect of the cluster's tidal field and flyby interactions. We find a significant synergy between both processes, where merger rates increase by a factor of $\sim 10-30$ compared to models in which only flybys or tides are taken into account.
This synergy is a consequence of the persistent tides-driven eccentricity excitation that is enhanced by the gradual diffusion of $j_z$  driven by flybys. The merger efficiency peaks when the diffusion rate is $\sim10-100$ slower than the tides-driven torquing. Added to this synergy, we also find that the gradual softening of the binary can lift the relativistic quenching of initially tight binaries, otherwise unable to reach extreme eccentricities, and thus expanding the available phase-space for mergers. Cumulatively, we conclude that despite the gradual softening of binaries due to flybys, these greatly enhance their merger rates in galaxy centers by promoting the tidal field-driven eccentricity excitation.


\vspace{0.75cm}

\end{abstract}

\section{Introduction}
\label{sec:introduction}

%Introduce GCs + SMBHs

Most nearby galaxies are known to harbor very dense stellar clusters at their core \citep{Neumnayer2011,Turner2012,Georgiev2014}. These nuclear star clusters (NSCs) tend to have masses of roughly $10^5-10^8M_\odot$, a small effective radii of only a few parsecs, and usually have a supermassive black hole (SMBH) residing at their center. This is the case for our own galaxy, containing an NSC with a mass of $\sim3\times10^7M_\odot$ \citep{Schodel2014,Schodel2017} and the central SMBH SgrA$^*$ with a mass of about $4\times10^6M_\odot$ \citep{Ghez2005,Gillessen2011}. This makes the galactic center (GC) a very dense and dynamical environment, which makes it an ideal laboratory to test several astrophysical phenomena involving stellar dynamics, general relativity (GR), and gravitational waves (GWs).

%(GW sources)
In the last decade there has been an increasing interest in gravitational-wave astrophysics, due to GW detections from the LIGO and VIRGO interferometers. Compact object binary mergers are known to be one of the main GW emitting events \citep{Abbott2016,Abbott2017}, therefore recent studies have focused on the dynamical mechanisms that lead to these mergers to better understand the sources of GWs. Particularly in GCs, one of the most important physical processes leading up to mergers is the torquing of binary's orbits produced by the SMBH, process known as the von Zeipel-Lidov-Kozai (ZLK) mechanism \citep{AntoniniPerets2012,Stephan2019,Hoang2019, Petrovich2017}. More generally, we can consider the tidal field from both the SMBH and the cluster which produces a behaviour similar to the ZLK mechanism \citep{Petrovich2017, hamilton2019_ApJL_mergers,BubPetrovich2020,Hamilton2021}.

The strong tidal field due to the SMBH and NSC potentials has demonstrated to play a key role in the evolution of binaries in the GC, as it efficiently torques the orbits of these binaries resulting in near unity eccentricities that lead to mergers. Recent efforts have been made in accounting for different tidal field-driven physical processes in binary evolution such as cluster triaxiality \citep{BubPetrovich2020} and relativistic phase space diffusion \citep{Hamilton2023}, however none has focused on combining these processes with others such as close encounters with field stars (flybys). Although the dynamics of wide binaries when under the influence of tidal fields \citep{Petrovich2017,Hamilton2019b,Rasskazov2023,Modak2023} and flybys \citep{CollinsSari2008,Leigh2017,Michaely2020,HamiltonModak2023} have been studied separately, the potential synergy between both effects has yet to be investigated in more depth.

Due to the high velocity dispersion $\sigma$ in the GC, most binaries in the GC are soft (e.g., $\sigma$ is larger than their Keplerian velocity) \citep{stephan2016,Alexander2014,Rose2020,Gautam2024}. One single encounter can potentially reset a binary's orbit, changing its orientation and even pushing it towards extreme eccentricities. The semi-major axis of these binaries will also tend to increase due to these encounters \citep{Heggie,Hillsa,Hillsb}. This orbit widening limits the efficiency for tidal field driven mergers, and previous works have simply prescribed it as a limiting evaporation timescale $t_\text{evap}$ (e.g., \citealt{stephan2016,Petrovich2017,Hoang2018}), using $t_\text{evap}$ from \cite{BinneyTremaine2008}. However, beyond the orbit softening, flyby interactions can produce a random walk in the binary's angular momentum $\vec{j}$ which may lead to mergers as studied by \cite{Michaely2020}. A study looking into the effect from flybys in the cluster-driven dynamics is largely missing in this context (GCs) and the focus of our work.

%(flyby=cluster in oort cloud)
The presented dynamics is reminiscent of that from wide binaries in general, including comets in the Oort Cloud. In particular, \citep{Heisler1986} studied how the galactic tidal field may torque the orbit of binaries and found that flybys can contribute to the diffusion in $j$, becoming a considerable contribution to the loss rate of comets in the Oort Cloud. In contrast, as we show in this paper, binaries in GCs have slow diffusion rates compared to the torquing rates that drastically promote the merger fraction. 

%(It has proven useful to study several different astrophysical problems, ranging from compact object binaries in dense stellar clusters to comets in the Oort Cloud) \citep{Heisler1986}. %These binaries are rarely isolated; they tend to be immersed in dynamically active environments such as stellar clusters, which can play an important role in their orbital evolution.

%(We do this and organized as)
In this paper we study the binary evolution due to both stellar flybys and tidal fields, proving that by combining both mechanisms can drive these binaries in galactic centers to extreme eccentricities much more efficiently than when these processes are considered separately. 
The paper is organised as follows. We introduce the basic dynamics for binaries in tidal fields in section \S\ref{sec:dynamical evolution in tf} and discuss the specific case of the GC including flybys in section \S\ref{sec:GC}. We present our results from numerical simulations in section \S\ref{sec:simulations} and we discuss the synergy found between tidal fields and flybys in \S\ref{sec:tides vs flybys}, along with the effects on relativistic quenching in section \S\ref{sec:GR quenching}. Finally, we present an overall discussion in section \S\ref{sec:discussion} and summarize and present the main conclusions in section \S\ref{sec:conclusions}. %In Appendix \ref{ap:IA} we show the derivations for the impulse approximation used to make the calculations presented in section \S\ref{sec:tides vs flybys}.

\section{Dynamical evolution due to tidal fields}
\label{sec:dynamical evolution in tf}
As binaries are usually immersed in stellar clusters, an important aspect of their dynamical evolution will be driven by the smooth tidal field of said cluster. These tidal fields drive secular oscillations in a binary’s eccentricity and inclination, similar to the ZLK mechanism. In this section we model this evolution mathematically, following the work done by \cite{Hamilton2019b,Hamilton2019a}. We will refer to the compact object binary system as the \emph{inner} binary, while the orbit of this system around the central SMBH will be referred to as the \emph{outer} binary.

\subsection{Equations of motion}
\label{subsec:equations of motion}
When studying orbits in axisymmetric potentials, the evolution of the inner binary can be approximated by averaging the tidal tensors over several periods of the outer binary. This is known as ``torus-averaging''; when the outer orbit of the binary is not closed, it will trace a non-repeating path around the cluster which will, after many orbits, densely fill an axisymmetric torus. For any function that follows the outer orbit of the binary, a weighted volume average over this torus can be used instead of a time average.

As shown by \cite{Hamilton2019b}, in this torus-averaged limit the tidal tensor of a given potential $\Phi$ only has diagonal terms, and $\langle\Phi_{xx}\rangle=\langle\Phi_{yy}\rangle$. Here it is useful to introduce Milankovitch's eccentricity and dimensionless angular momentum vectors $\{\Vec{e},\Vec{j}\}$, Milankovitch's eccentricity and dimensionless angular momentum vectors (e.g., see \citealt{Book_Tremaine2023}), in order to simplify notation. For a smooth potential $\Phi$ that changes over scales that are much larger than that of the inner binary's semi-major axis (expanded up to the quadrupole tidal field), this results in the torus-averaged potential
\begin{equation}
    \langle\Phi\rangle =\frac{3a\iB^2}{2} \langle\Phi_{zz}+\Phi_{xx}\rangle\left[\frac{1}{2}\Gamma(5e_z^2-j_z^2)+\f{1}{4}e^2(1-5\Gamma)\right],
\end{equation}
where
\begin{equation}
    \Gamma \equiv \frac{\langle\Phi_{zz}-\Phi_{xx}\rangle}{3\langle\Phi_{zz}+\Phi_{xx}\rangle}.
\end{equation}

From this potential, a doubly-averaged Hamiltonian can be written in terms of $\Gamma$  as $H=H_K+H_1$ where $H_K $ is the Keplerian term and the dimensionless perturbing term is given by \citep{Hamilton2019a}:
\begin{equation}
\label{eq:H1_2}
\begin{split}
    H_1=2\Bigl[1-\frac{3}{2}e^2(5\Gamma-1)-3\Gamma(\Vec{j}\cdot\hat{j}\oB)^2+15\Gamma(\Vec{e}\cdot\hat{j}\oB)^2\Bigr].
\end{split}
\end{equation}
 The associated dimensionless time is $\tau=t/\tau_\text{sec}$, where  $\tau_\text{sec}$ is the secular timescale given by
\begin{equation}
\label{eq:tau_sec1}
\tau_\text{sec}^{-1} = \frac{3a\iB^{3/2}}{2\sqrt{GM\bin}}\langle\Phi_{zz}+\Phi_{xx}\rangle.
\end{equation}

The equations of motion for the binary that we use throughout this work can be written in a vectorial form as:
\begin{equation}
\label{eq:e}
\begin{split}
    \frac{d\Vec{e}}{d\tau}&=\frac{1}{2}(5\Gamma-1)(\Vec{j}\times\Vec{e})-5\Gamma(\Vec{e}\cdot\hat{j}\oB)(\Vec{j}\times\hat{j}\oB)\\
    &\qquad+\Gamma(\Vec{j}\cdot\hat{j}\oB)(\Vec{e}\times\hat{j}\oB),\\
\end{split}
\end{equation}

\begin{equation}
\label{eq:j}
\begin{split}
    \frac{d\Vec{j}}{d\tau}&=\Gamma(\Vec{j}\cdot\hat{j}\oB)(\Vec{j}\times\hat{j}\oB)-5\Gamma(\Vec{e}\cdot\hat{j}\oB)(\Vec{e}\times\hat{j}\oB),
\end{split}
\end{equation}
where we shall assume that $\hat{j}\oB$ remains fixed throughout the evolution. The latter approximation assumes that either the tidal field is spherically symmetric or the outer orbit lies in the meridional plane of an axisymmetric field \citep{Petrovich2017}.

As discussed by \citet{Hamilton2019a,Hamilton2019b}, $\Gamma$ determines the phase-space structure of the Hamiltonian, and hence is a useful parameter to classify different dynamical regimes for binary evolution. For example, $\Gamma=1$ when the potential is Keplerian (only a SMBH at the center), and hence the Hamiltonian in equation (\ref{eq:H1_2}) reduces to the quadrupole ZLK Hamiltonian. The secular timescale reduces to the well-known ZLK timescale (e.g., \citealt{holman1997}):
%\begin{align}
\ba
\label{eq:tau_ZLK}
    \tau_{\rm ZLK}&=&\frac{4a^3\oB(1-e\oB^2)^{3/2}M^{1/2}\bin}{3a\iB^{3/2}M\BH G^{1/2}}\nonumber\\
    &\simeq& 3\times 10^{5}\left(\frac{M_{\rm bin}}{10M_\odot}\right)\left(\frac{M\BH}{4\times10^6M_\odot}\right)^{-1} \nonumber \\
    &\times&\left(\frac{a\iB}{25\rm au}\right)^{-3/2}\left(\frac{a\oB(1-e^2\oB)^{1/2}}{0.3\rm pc}\right)^3\text{year}.
\ea
%\end{align}
We note that the ZLK Hamiltonian assumes the test particle approximation and sets $e\oB=0$, so that the quadrupole level expansion is a good approximation of the secular dynamics. If, instead, the outer orbit is eccentric, the $z$-component of the angular momentum is no longer conserved, and hence the octupole expansion is required \citep{naoz_Kozai_review}. This expansion allows for qualitatively different behavior, where extremely high eccentricities for the inner orbit can be achieved along with orbital flipping \citep{Naoz2011,Naoz2013, Hoang2018}. 

Similarly, we can obtain the Hamiltonian for a binary evolving in a thin Galactic disk by replacing $\Gamma=1/3$ (see discussion in \S 7.1), while realistic spherical potentials can be reproduced with $0\leqslant\Gamma\leqslant1$ as shown by \cite{Hamilton2019b}.

An important property of the secular Hamiltonian in Equation (\ref{eq:H1_2}) is its axissymmetry 
(i.e., invariability to rotations around $\hat{j}\oB$ ), which implies the $z$-component of the angular momentum $\propto\sqrt{a\iB(1-e\iB^2)}\cos i$ with $\cos i\iB=\hat{j}\iB\cdot\hat{j}\oB$ is conserved. Thus, the Hamiltonian is integrable and we can express the maximum eccentricity due to tidal fields as \citep{Hamilton2019b}: 
\begin{equation}
\label{eq:e_max}
    e_\text{max} = \sqrt{1-\frac{10\Gamma\cos^2{i_0}}{5\Gamma+1}},
\end{equation}
where $i_0$ is the initial initial when $e_0^2$ is nearly 0.

Finally, in addition to the equations of motion  (\ref{eq:e}) and (\ref{eq:j}), we add an extra term in Equation  (\ref{eq:e}) to account for the relativistic precession given by
\begin{equation}
\label{eq:GR precession}
    \f{d\Vec{e}}{d t}\Big|\GR=\f{\dot{\omega}\GR}{(1-e^2)^{3/2}}\Vec{j}\times\Vec{e},
\end{equation}
where 
\begin{equation}
    \dot{\omega}\GR= \f{3G^{3/2}M\bin^{3/2}}{a\iB^{5/2}c^2}.
\end{equation}

%where $\epsilon\GR$ is the ratio between the secular timescale and the GR precession timescale, which parametrizes the importance of relativistic precession as \citep{Petrovich2017}

%\begin{equation}
%    \epsilon\GR\equiv \f{\tau_\text{sec}}{\tau\GR}%=\f{4GM\bin^2a\oB^3(1-e\oB^2)^{3/2}}{c^2a\iB^4M\BH}
%\end{equation}

%where the larger the value of $\epsilon\GR$, the faster the relativistic precession.

\subsection{Merger conditions}
\label{subsec:merger}

A major application we consider in the galactic center is the evolution and potential merger of compact object binaries. When the two components of these binaries come very close together, they lose energy by GW radiation. The merger timescale for this gravitational radiation is defined by \cite{Peters1964} as
\begin{equation}
    \label{eq:tau_GW}
    \tau_\text{GW} = \frac{3}{85}\l(\frac{a\iB^4c^5}{G^3m_1m_2M\bin}\r)(1-e\iB)^{7/2},
\end{equation}

where $m_1$ and $m_2$ If a binary's orbit shrinks significantly in one secular cycle, then a rapid merger will take place. For this to happen, the timescale for the change in the semi-major axis of the binary due to GW losses ($\tau_\text{GW}$) must be shorter than the timescale for the change in the pericentric distance $a\iB(1-e\iB)$ due to tidal fields, which, at high eccentricities, is $\simeq\sqrt{1-e^2}\tau_{\rm sec}$. This results in the condition 
\begin{equation}
\label{eq:e_merger}
    1-e_\text{merger}\lesssim 3\times10^{-5}\l(\frac{a\oB(1-e\oB^2)^{1/2}}{0.1\text{pc}}\r),
\end{equation}
for binaries with $M\bin=10M_\odot$ initially at $a\iB=10$au when only ZLK cycles are considered (only SMBH). A similar expression can be obtained for a more general cluster potential (arbitrary $\Gamma$; see Equation \ref{eq:tau_sec} for a secular timescale adding a cluster with a Hernquist potential).

In order for BH (stellar) binaries to reach extremely high eccentricities of  $1-e_{\text{max}}\lesssim10^{-5}$ ($1-e_{\text{max}}\lesssim10^{-3}$), we require that $\dot{\omega}\GR\tau_\mathrm{sec}\lesssim0.01$ ($\dot{\omega}\GR\tau_\mathrm{sec}\lesssim0.1$)  as shown by \cite{Petrovich2017}. In general, we will define extreme eccentricities as $1-e_{\text{max}}\lesssim10^{-4}$ following equation (\ref{eq:e_merger}), hence binaries that achieve these values will be considered as possible mergers.

\section{Dynamics in the Galactic center}
\label{sec:GC}

In the case of a spherical $\gamma$-family cluster with a distance scale $s$ and with a central SMBH in which we have a circular orbit, $\Gamma$ becomes 
\begin{equation}
    \Gamma = \f{3M\BH+M\cl a\oB^{3-\gamma}(a\oB+s)^{\gamma-4}(3a\oB+s\gamma)}{3[M\BH+M\cl a\oB^{3-\gamma}(a\oB+s)^{\gamma-4}(a\oB+s(4-\gamma))]},
\end{equation}
where we have considered the tidal averaged tidal tensors $\langle\Phi_{xx}\rangle$ and $\langle\Phi_{zz}\rangle$ as derived by \cite{BubPetrovich2020}.

In the simplified case that we consider a spherical Hernquist potential with $\gamma=1$ to model the center of our galaxy, this reduces to
\begin{equation}
    \Gamma = \f{3M\BH+M\cl a\oB^{2}(a\oB+s)^{-3}(3a\oB+s)}{3[M\BH+M\cl a\oB^{2}(a\oB+s)^{-3}(a\oB+3s)]},
\end{equation}
where $s$ is the scale radius of the potential.

For small values of $a\oB/s$ we have $\Gamma\approx1$, where the SMBH dominates over the cluster and hence the dynamical evolution of the binary reduces to the ZLK mechanism as mentioned in section \S\ref{subsec:equations of motion}. The same happens when $a\oB/ s$ becomes very large; the cluster and SMBH in this limit behave as a single body so that as we get further away from the central SMBH, $\Gamma\rightarrow1$. 

The secular timescale in these kinds of environments then becomes \citep{BubPetrovich2020}:
\begin{equation}
\begin{split}
    \label{eq:tau_sec}
    \tau_{\text{sec}}^{-1}=&\f{3a\iB^{3/2}}{2\sqrt{GM\bin}}\Bigg(\frac{GM\BH}{2a\oB^3}\\
    &+\frac{GM\cl}{a\oB^\gamma(a\oB+s)^{3-\gamma}}\l[2-\frac{3a\oB+s\gamma}{2(a\oB+s)}\r]\Bigg).
\end{split}
\end{equation}
%with which equation (\ref{eq:GR precession}) becomes

%\begin{equation}
%\label{eq:GR precession2}
%5    \f{d\Vec{e}}{d\tau}\Big|\GR=\f{\epsilon\GR}{(1-%e^2)^{3/2}}\Vec{j}\times\Vec{e}
%\end{equation}

\begin{figure*}
\centering
\includegraphics[width=\linewidth]{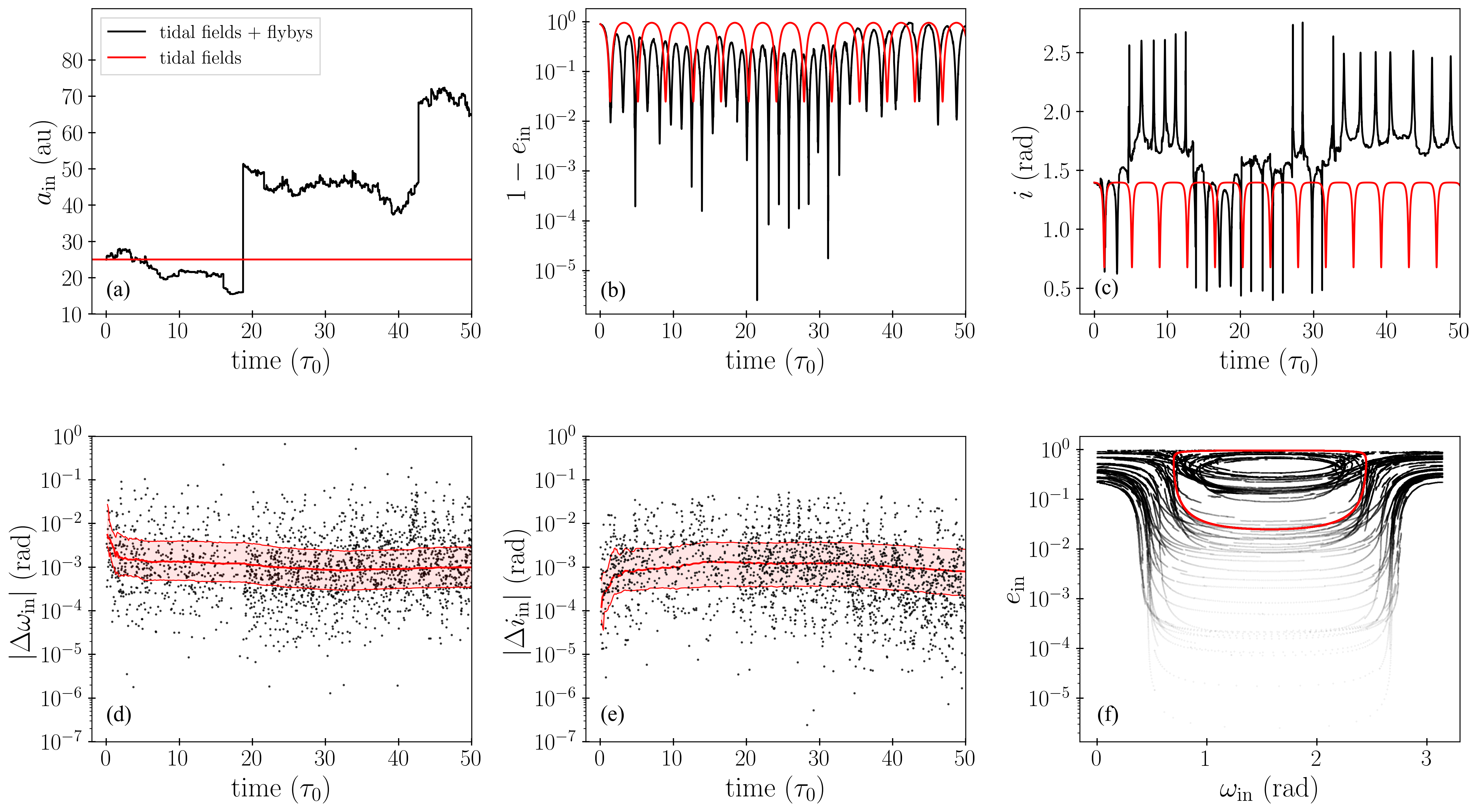}
\caption{Full evolution of a binary due to both tidal fields and flyby effects. The initial parameters are $a\oB=0.3$ pc,  $a\iB=25$ AU, $e\iB=0.1$, $\omega\iB=\pi/4$, $\Omega\iB=0$, $i\iB=80$°. Each stellar component of the binary has a mass of $5M_\odot$, and we consider the fiducial scenario of $M\BH=4\times10^6M_\odot$. The red curves represent the evolution in a flyby-free scenario, simply by integrating Equations (\ref{eq:e}) and (\ref{eq:j}). In panels (d) and (e), the shaded region contains the median and both upper and lower quartiles.}
\label{fig:complete-evolution}
\end{figure*}

\subsection{Fly-bys}
\label{sec:flybys}
The main focus of this work is to study the additional effect produced by flybys on the evolution of binaries in tidal fields. Due to the high velocity dispersion in the GC, the binaries residing there are mostly soft ($v_\text{orb}<\sigma$, with $\sigma$ the velocity dispersion of the cluster), hence we expect flyby encounters to produce a systematic drift in the semi-major axis of the binary. This is a consequence of the Heggie-Hills law, derived using thermodynamic arguments: "hard binaries tend to harden, whereas soft binaries tend to soften" \citep{Hillsa,Hillsb,Heggie}.

%%%%%%%%%%%%
In order to classify binaries between hard and soft, we can write the velocity dispersion in the galactic center as according to \cite{Kocsis2011} 

\begin{equation}
\label{eq:sigma}
    \sigma(r)=    \begin{cases}
         280\text{kms}^{-1}\sqrt{0.1\text{pc}/r}\sqrt{1-0.035(r/0.1\text{pc})^{2.2}},\\
          \phantom{ 280\text{kms}^{-1}\sqrt{0.1\text{pc}/r}\sqrt{1-0.03}}\text{if }r<0.22\text{pc}\\
       250\text{kms}^{-1}\sqrt{0.1\text{pc}/r}, \phantom{ \sqrt{01\text{pc}/r}} \text{if }  r >0.22\text{pc}
    \end{cases}
\end{equation}

The hard-soft boundary then is given by $v_{\rm{kep}}=\sigma$. For example, if we consider a twin binary with two $5M_\odot$ components at $a\oB=0.3$pc , this boundary would be given by $a\iB=0.2$au. If we increase the mass of the binary component this limit will occur at a larger value; for $10M_\odot$ components it occurs at $a\iB=0.4$au, and for $50M_\odot$ at $a\iB=2.1$au. These continue to be relatively small values, further reinforcing that most binaries in the galactic center can be classified as soft.

The time between flyby encounters

\begin{equation}
\label{eq:enc_time}
    t_{\rm enc}=(n_\star\sigma_{p} v_p)^{-1},
\end{equation}
where $\sigma_{p} = \pi b_\text{max}^2$ is the geometric cross section considered for the flybys and $n_\star$ is the stellar number density of the environment. For the galactic center it is considered as $n_\star\approx 10^6\text{pc}^{-3}$ \citep{HamersTremaine2017} and $n_\star\approx 0.1\text{pc}^{-3}$ for the solar neighbourhood. It can be more accuratelly modelled by \citep{Kocsis2011}

\begin{equation}
    n_\star(r)=\frac{2.8\times10^6\text{M}_\odot\text{pc}^{-3}}{\langle m_\star \rangle}\l(\frac{r}{0.22\text{pc}}\r)^{-\gamma},
\end{equation}
with $\gamma=1.2$ inside 0.22pc and $\gamma=1.75$ outside 0.22pc. We take $\langle m_\star\rangle\approx1M_\odot$ as the average stellar mass in the GC; \cite{Alexander2013} presents equations that result in $\langle m_\star\rangle$ ranging from $1M_\odot$ to $2M_\odot$, therefore to simplify calculations we take all flybys to have a solar mass.

\section{Numerical simulations}
\label{sec:simulations}

Because the interaction time due to a flyby (order $a\iB/\sigma$, $\sim 0.2$ years) is much shorter than the secular timescale, we evolve an individual binary using the secular equation of motions in (\ref{eq:e}) and (\ref{eq:j}) subject to discrete instances where a flyby takes place, modifying the orbital elements of the binary, and continue the secular evolution until the next flyby takes place. In  other words, we switch back and forth between tidal fields and flybys.

The flybys are modelled using  direct N-body integrations with REBOUND \citep{ReinLiu2012}, with an encounter rate as specified in equation (\ref{eq:enc_time}) defining $b_\text{max}=5a\iB$ to avoid considering flybys that are too distant. This definition allows us to stay close to the regime in which the impulse approximation is valid, which will prove to be relevant in section \S\ref{sec:tides vs flybys}. We repeated our fiducial numerical experiments in which we doubled $b_\text{max}$ to $10a\iB$, and found that the outcome was statically similar to those with $b_\text{max}=5a\iB$, plus the fact that a higher value for $b_\text{max}$ implies a higher frequency for flyby encounters, which is much more computationally expensive to simulate.

We establish the velocity of our flybys to be constant at $v_p=200$ kms$^{-1}$, considering that within 1 pc of the center of our galaxy $\sigma$ varies roughly between 100 and 300 kms$^{-1}$ as according to equation (\ref{eq:sigma}). These simplifications allow for controlled simulations, which makes comparisons with analytical estimates such as the ones made in section \S\ref{sec:tides vs flybys} easier. 

Figure \ref{fig:complete-evolution} shows the complete dynamical evolution of an individual binary with $a\oB=0.3$pc. The red curves show the evolution of an identical binary in a flyby-free scenario, where we only get tidal field driven evolution. It is quite clear by comparing these curves with our results that the inclusion of flybys in our model results in a more dramatic evolution which involves extreme eccentricities, orbital flips and jumps between the $e\iB -\omega\iB$ phase-space curves. 

The individual changes in $a\iB$ presented in panel (a) follow a random walk, but show an overall tendency of making the binary wider. This is expected due to the Heggie-Hills law, as mentioned earlier. We can also easily see in panel (b) that by only taking into account tidal field driven evolution, such high eccentricities would never have been achieved by this particular binary. However, this becomes possible when including flybys in the model, which allows the binary to reach values of $1-e\iB\approx10^{-5}$. As discussed in section \S\ref{subsec:merger}, with these extreme eccentricities a rapid merger can be produced due to GW tidal captures.

It is also interesting to note that in panel (c) we can see orbital flips, in which the orbit switches between prograde and retrograde. Considering that the Hamiltonian shown in Equation (\ref{eq:H1_2}) is only of quadrupole order and assumes $e\oB=0$, these flips should not be possible as $j_z=\sqrt{1-e\iB^2}\cos i\iB$ is fixed (e.g., \citealt{naoz_Kozai_review}). However, as flybys modify both $e\iB$ and $i\iB$ driving a slow random walk in $j_z$ (see also the examples in Figure \ref{fig:R_cases}), the orbit is allowed to flip, a phenomenon observed during the high eccentricity episodes.

%It is relevant to note that this hamiltonian alone doesn't allow for very dramatic dynamical behaviour. However in the case of an eccentric outer orbit, for example, the z-component of the angular momentum is no longer conserved and hence the octupole expansion is required. This leads to qualitatively different behaviour, where extremely high eccentricities for the inner orbit can be achieved  \citep{Naoz2011,Naoz2013,Teyssandier2013} and the orbit can also flip between prograde and retrograde \citep{Naoz2011,Naoz2013}.

Panels (d) and (e) show the variation in $\omega\iB$ and $i\iB$ product of each flyby encounter. We can see that both these variations in general are very small, however, along with the variations in eccentricity due to each encounter, they are enough to make the binary's orbit jump from one phase-space curve to another as is shown in panel (f). Since flybys not only affect the magnitude of the eccentricity but also the orientation of the orbit, we get changes both in $\omega\iB$ and in $e\iB$ which produce these `jumps' in the phase-space portrait. In particular, in a flyby-free scenario the orbit would librate around the fixed point $\omega\iB=\pi/2$, however by including flybys in our model the orbit generally circulates one as shown in panel (f).

\subsection{Population statistics}
\label{subsec:pop synthesis}

\begin{figure}
\centering
\includegraphics[width=\linewidth]{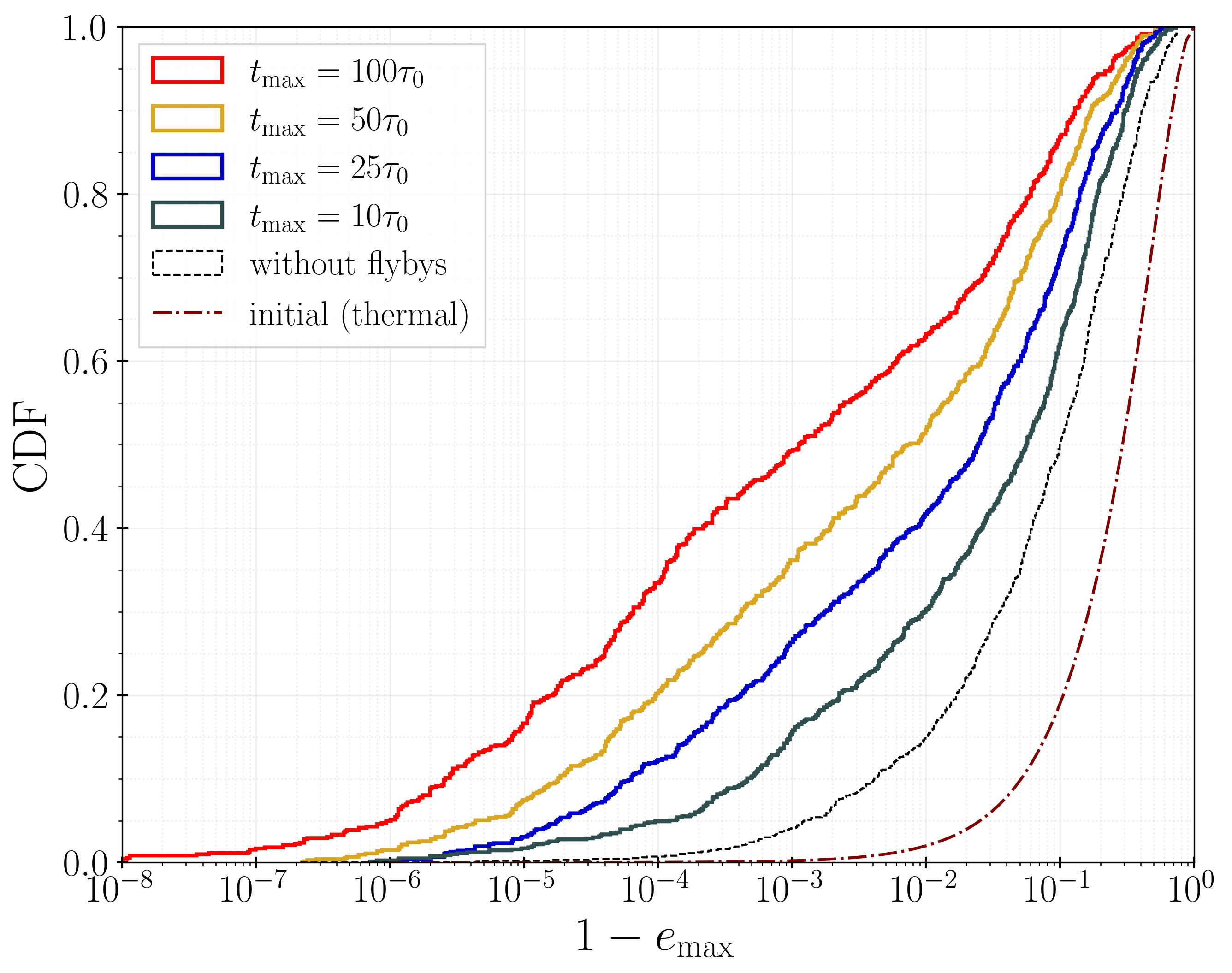}
\caption{CDFs of the maximum eccentricities reached in the galactic center after 10, 25, 50 and 100 initial secular timescales $\tau_0$, as labeled. The dashed histogram represents the CDF for a flyby free population, evolved for $100\tau_0$.}
\label{fig:cdf-rebound}
\end{figure}

In order to quantify the overall effects of tidal fields and flybys on binary evolution, we generated a series of simplified population synthesis models. It is important to note that all binaries within the population are independent from one another, hence we are not considering possible effects due to binary-binary interactions. We also consider a dynamical stability criterion, so as to not consider binaries that become extremely wide and hence unstable when \citep{Grishin2016}
\begin{equation}
\label{eq: stability}
    a\iB>0.25a\oB\l(\frac{M\bin}{M\BH
}\r)^{1/3}.
\end{equation}

In Figure \ref{fig:cdf-rebound} we show the CDF of the maximum eccentricities achieved by a population of $\sim700$ binaries in the galactic center at different stages of their evolution. The initial eccentricities of these binaries follow a thermal distribution such that $f(e)de=2ede$, and their initial inclinations are uniform such that $\cos{i_0}$ is also uniform between -1 and 1. The eccentricity of the outer orbits is set to 0. All binaries are twins with a total mass of $10M_\odot$ and are evolved for $100\tau_0$, $\tau_0$ being the initial secular timescale of each individual binary according to equation (\ref{eq:tau_sec}). We consider the fiducial scenario of our GC, where $M\BH=4\times10^6M_\odot$, $M\cl\approx8M\BH$, and $s=4$ pc. The initial values for both $a\iB$ and $a\oB$ are drawn from a uniform distribution, ranging between $15-50$au for $a\iB$ and $0.1-1$pc for $a\oB$. This means that the total population of binaries is initially soft, and that the velocity dispersion does not exceed $275$kms$^{-1}$ making our assumption of $v_p=200$kms$^{-1}$ appropriate.

After only a few secular timescales, the initial thermal eccentricity distribution is modified into a wider distribution, achieving higher eccentricities as expected from the tidal fields. We can also observe the cumulative nature of this type of dynamical evolution; the longer we let the binaries evolve, the higher the probability of them reaching extreme eccentricities. This is expected from a chaotic nature of the binary evolution depicted in Figure \ref{fig:complete-evolution}.

\begin{figure}
\centering
\includegraphics[width=\linewidth]{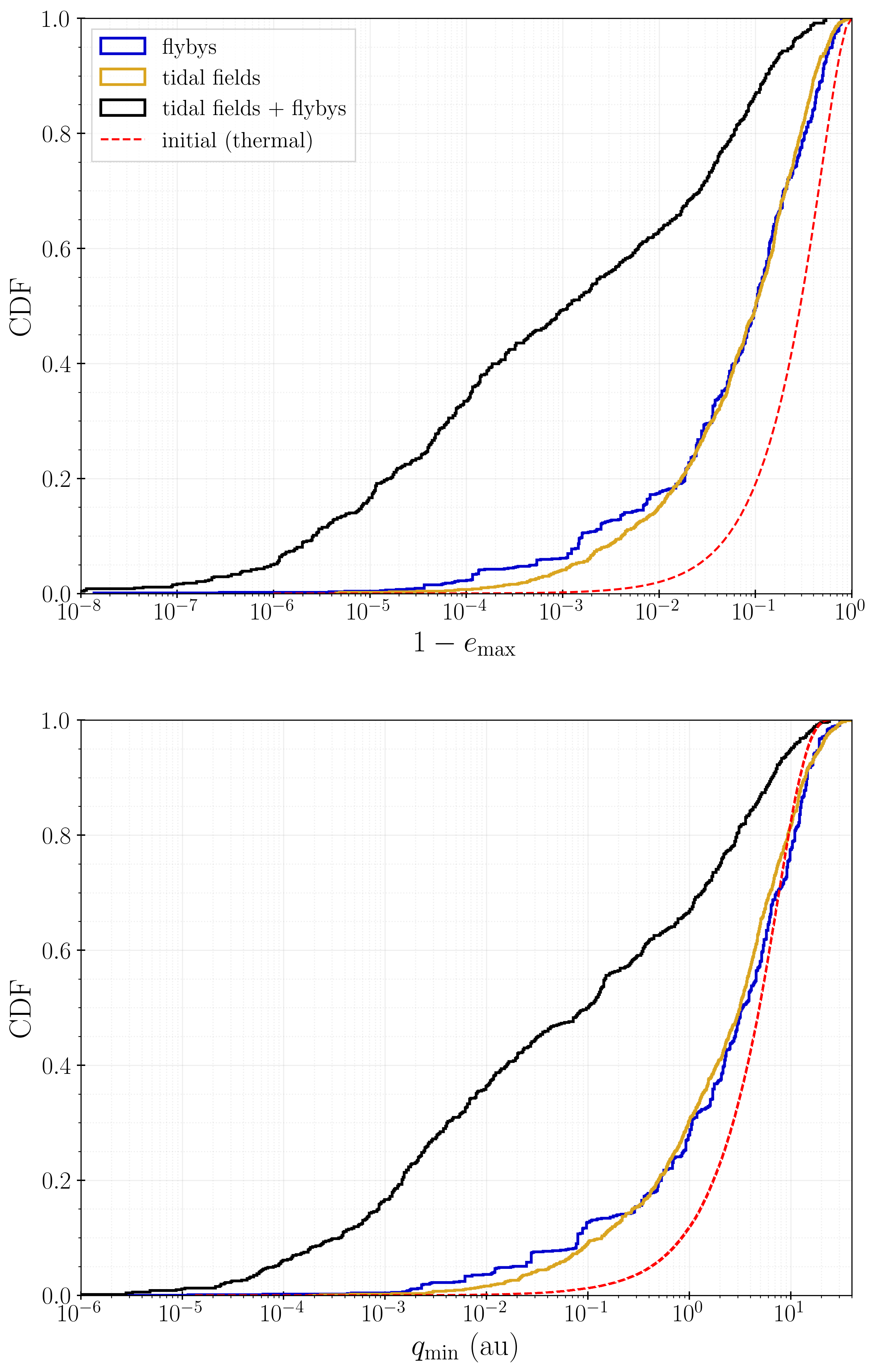}
\caption{CDFs of the maximum eccentricities and minimum pericenter distances $q\iB=a\iB(1-e\iB)$ reached in the galactic center after 100 initial secular timescales $\tau_0$ for different dynamical evolution mechanisms: tidal fields, flybys and a combination of both.}
\label{fig:cdf-all}
\end{figure}

\begin{figure}
\centering
\includegraphics[width=\linewidth]{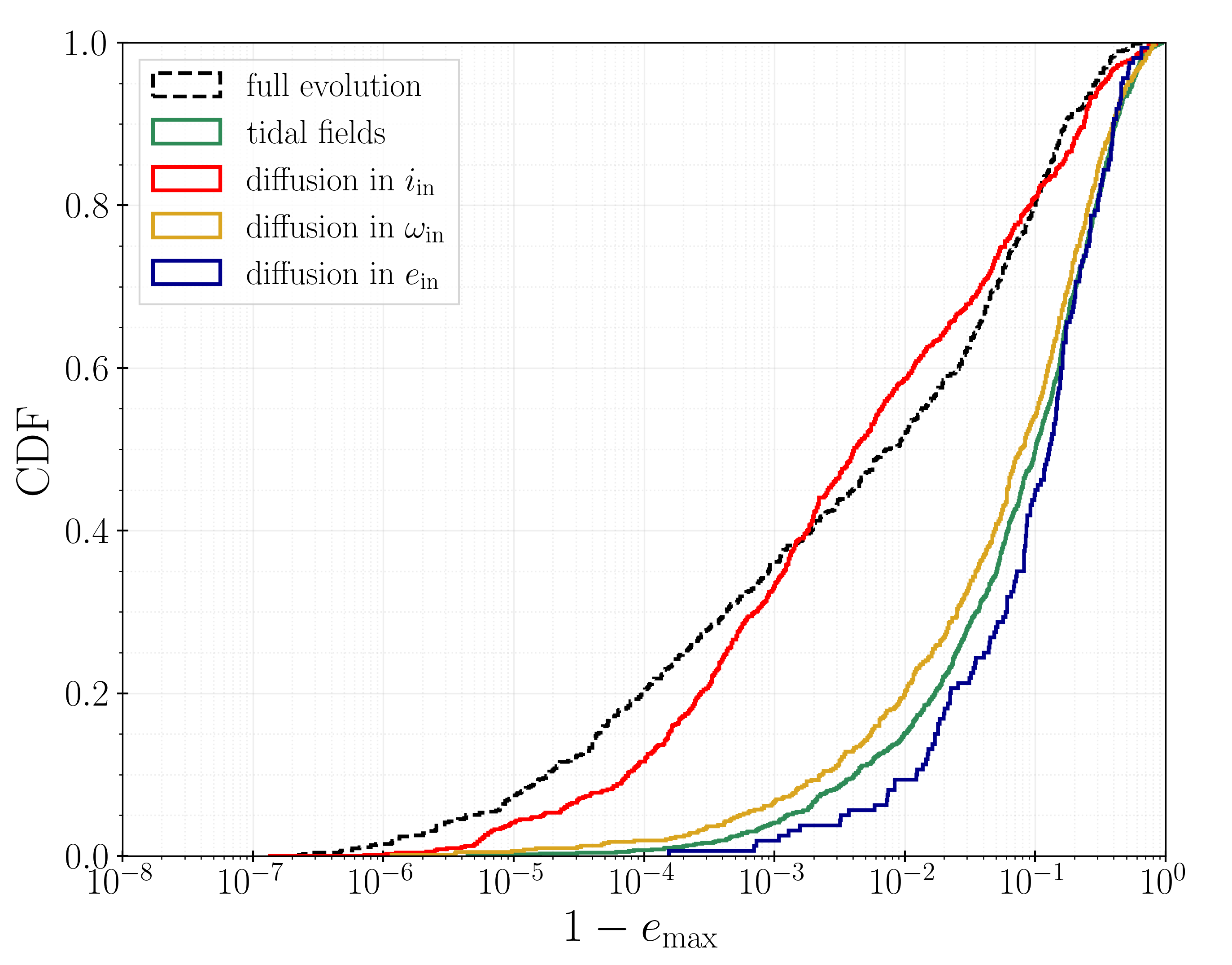}
\caption{CDFs of the maximum eccentricities achieved by diffusion in different orbital parameters, evolved for $50\tau_0$. The black dashed curve represents evolution considering diffusion in all the orbital parameters due to flybys; it is the same as the yellow curve shown in figure \ref{fig:cdf-rebound}.}
\label{fig:pop_diffusion}
\end{figure}

\begin{figure*}
\centering
\includegraphics[width=\linewidth]{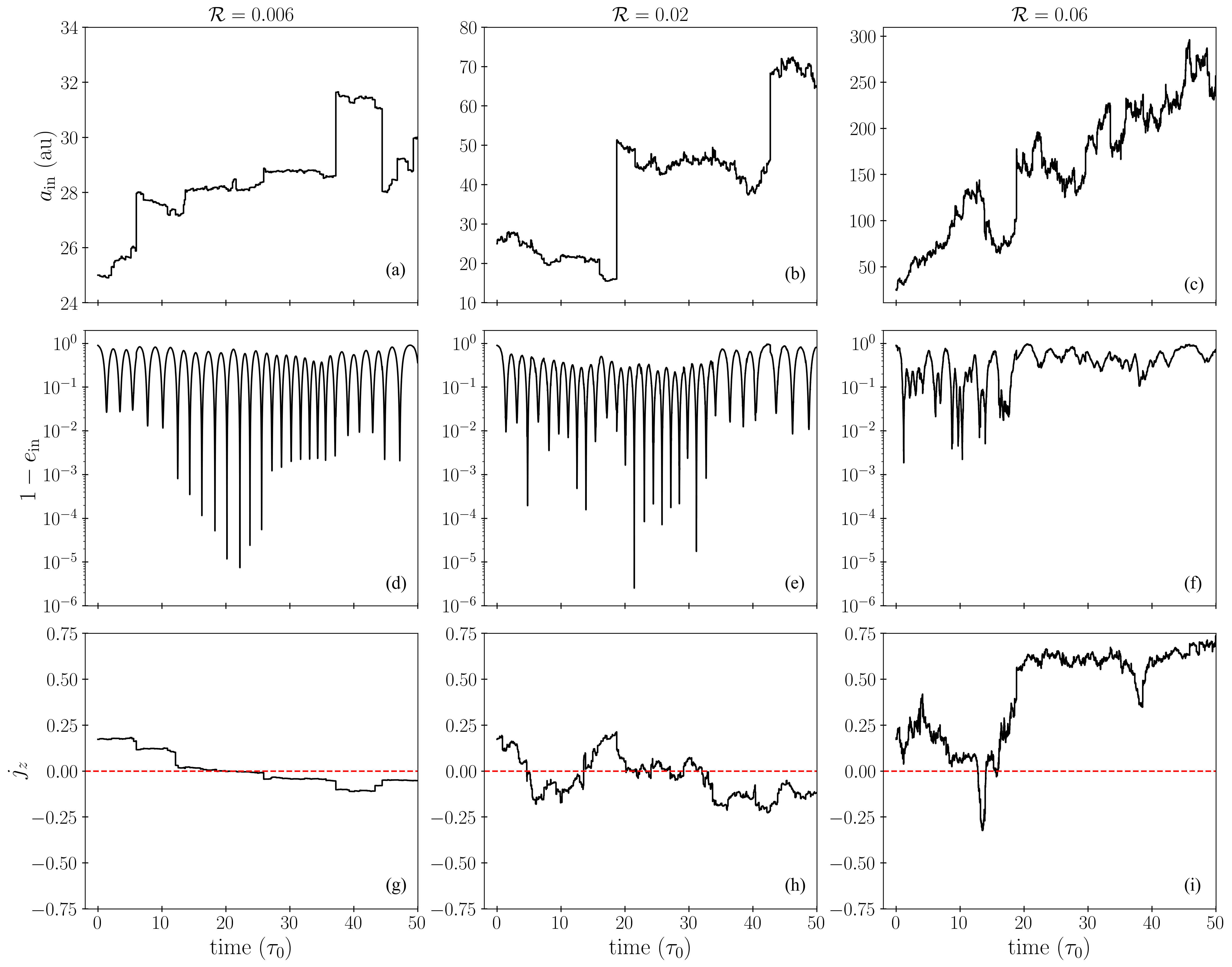}
\caption{Evolution of orbital parameters for three binaries with different values of $\mathcal{R}$, all with the same initial conditions of $a\iB=25$au, $a\oB=0.3$pc, $e\iB$=0.1 and $e\oB=0$. All were evolved for $50\tau_0$, $\tau_0$ being the initial secular timescale for each binary. Note the difference in the scale of the vertical axis in the top row; as $\mathcal{R}$ increases, $a\iB$ grows more drastically.}
\label{fig:R_cases}
\end{figure*}

In Figure \ref{fig:cdf-all} we can see the CDFs of the maximum eccentricities achieved after $100\tau_0$ by three different populations of binaries. One population was evolved considering only tidal field effects, the second considering only flybys and the third taking into account both effects (the same population as shown in figure \ref{fig:cdf-rebound}). The figure shows that tidal fields are the least efficient dynamical mechanism in producing extreme eccentricities; only about $\sim1\%$ of binaries evolved with tidal fields alone reach  $1-e_\text{max}<10^{-4}$. Flybys on the other hand prove to be slightly more effective, driving about $3\%$ of binaries in the population to said values. However, when combining both mechanisms a clear synergy is found, increasing this probability up to $\sim34\%$. 
For reference, the bottom panel shows the minimum pericenter distances for the same population of binaries. Although several binaries reach very high eccentricities, as was mentioned in section \S\ref{sec:flybys} these binaries will tend to get softer, which means that $a\iB$ and hence the pericentric distance $q=a\iB(1-e\iB)$ will grow as well. This is where the stability criterion mentioned previously in equation (\ref{eq: stability}) becomes important. As can be seen in the CDF, the extreme values of the eccentricities achieved seem to dominate over the increase in $a\iB$, producing a considerable amount of binaries with pericentric distances smaller than $\sim10^{-3}$au, equivalent to a merger timescale $\tau_{\rm GW}<3\times 10^5$ yrs (Eq. [\ref{eq:tau_GW}]). The latter may be shorter than the ZLK timescale in Equation (\ref{eq:tau_ZLK}) and, therefore, expected to lead to a merger within one secular cycle.
\section{Tides vs. flybys}
\label{sec:tides vs flybys}
\begin{figure}
\centering
\includegraphics[width=\linewidth]{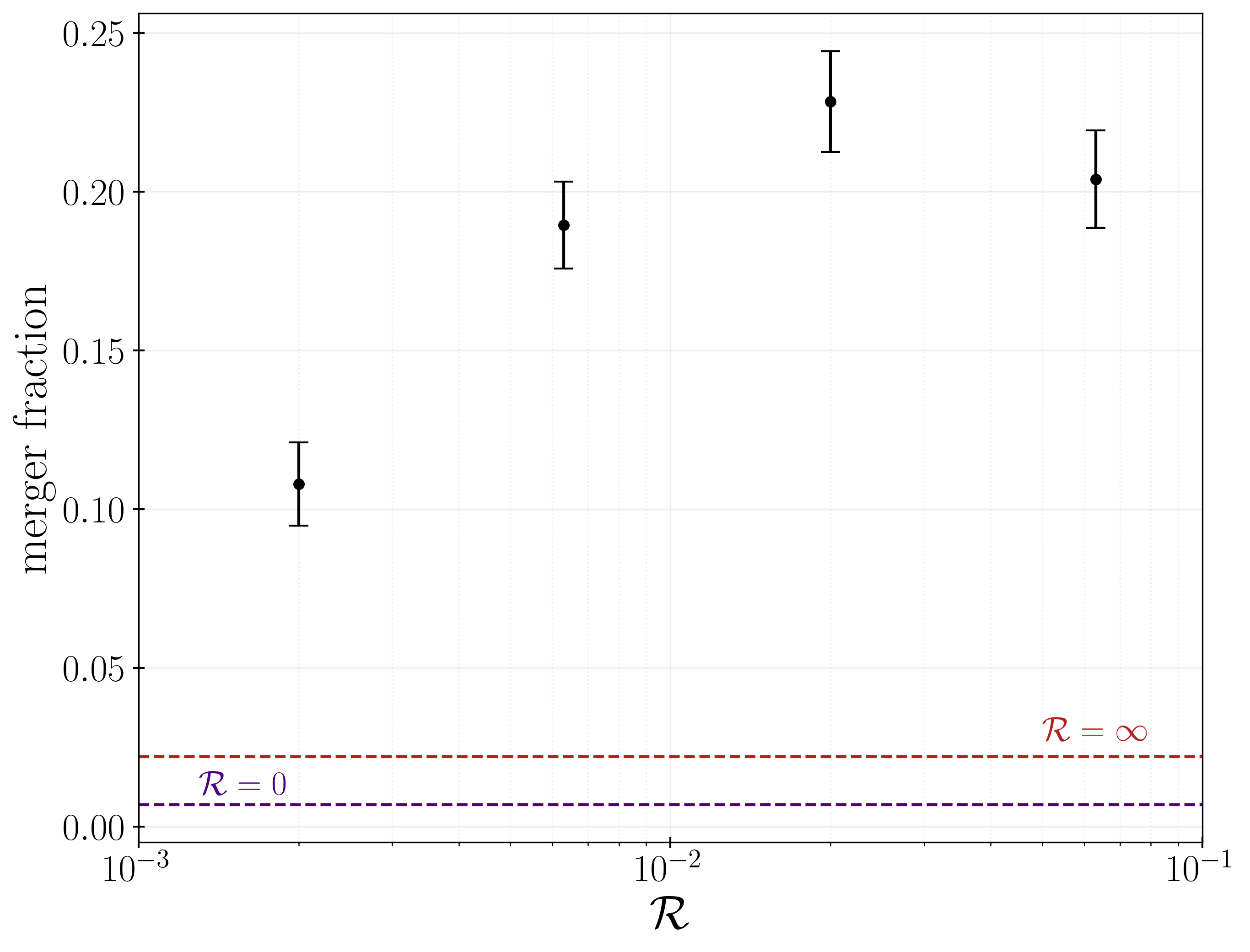}
\caption{Merger fraction as a function of the diffusion parameter $\mathcal{R}$. The dashed lines indicate the merger fractions for the limits $\mathcal{R}=0$ (tidal fields alone) and $\mathcal{R}=\infty$ (flybys alone). The error bars are evaluated assuming a Poisson distribution as $\sqrt{N}$, $N$ being the amount of mergers for each bin.}
\label{fig:merger_frac}
\end{figure}
\begin{figure}
\centering
\includegraphics[width=\linewidth]{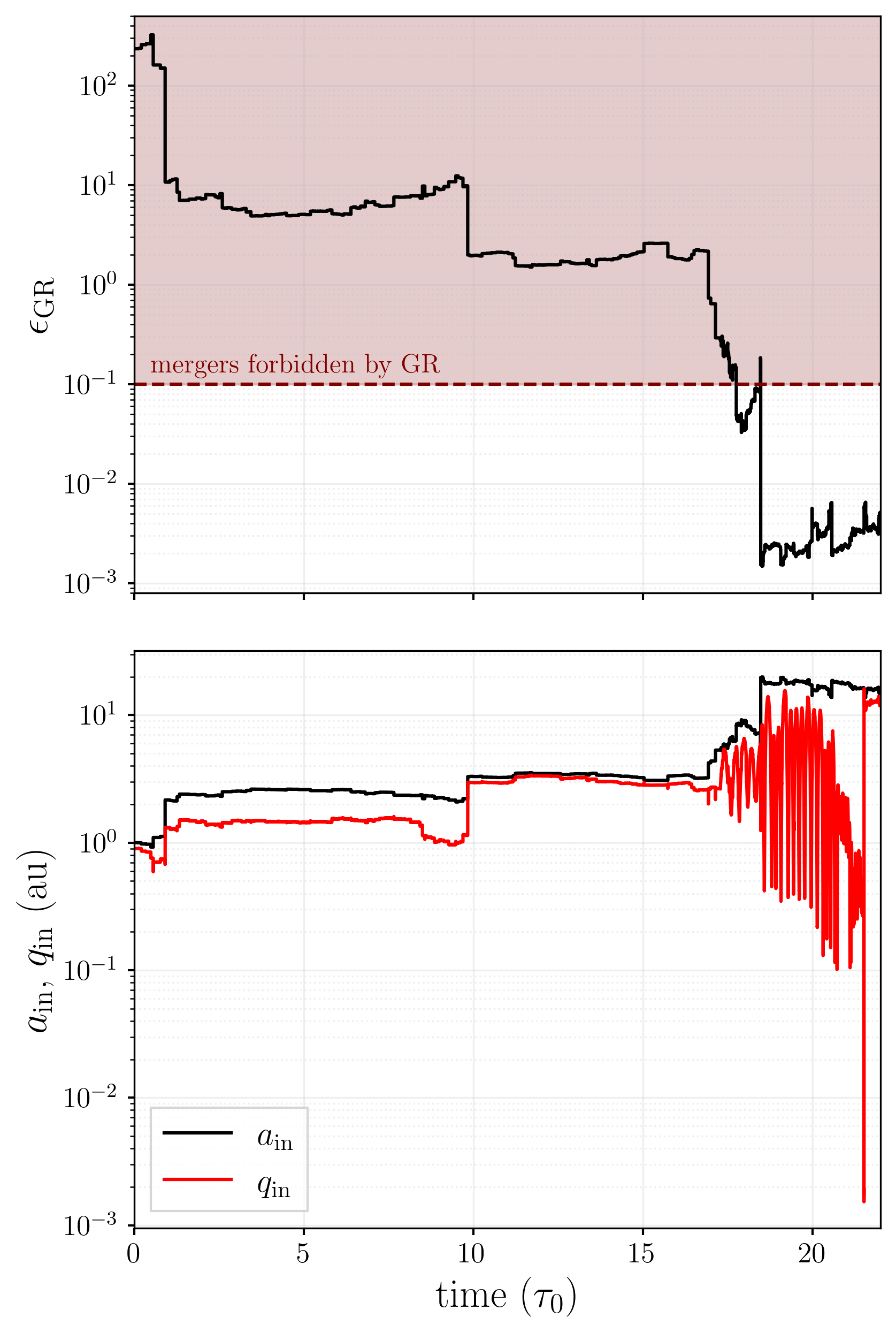}
\caption{Evolution of a binary that begins in a GR quenched regime in which mergers are forbidden (colored region). Flyby interactions allow this binary to become softer and move out of this regime, hence becoming a merger candidate. The top panel shows the evolution of $\epsilon_\text{GR}$ as according to equation (\ref{eq:GR precession}), and the bottom panel shows the evolution of $a\iB$ and $q\iB$.}
\label{fig:gr_case}
\end{figure}

\begin{figure*}
\centering
\includegraphics[width=\linewidth]{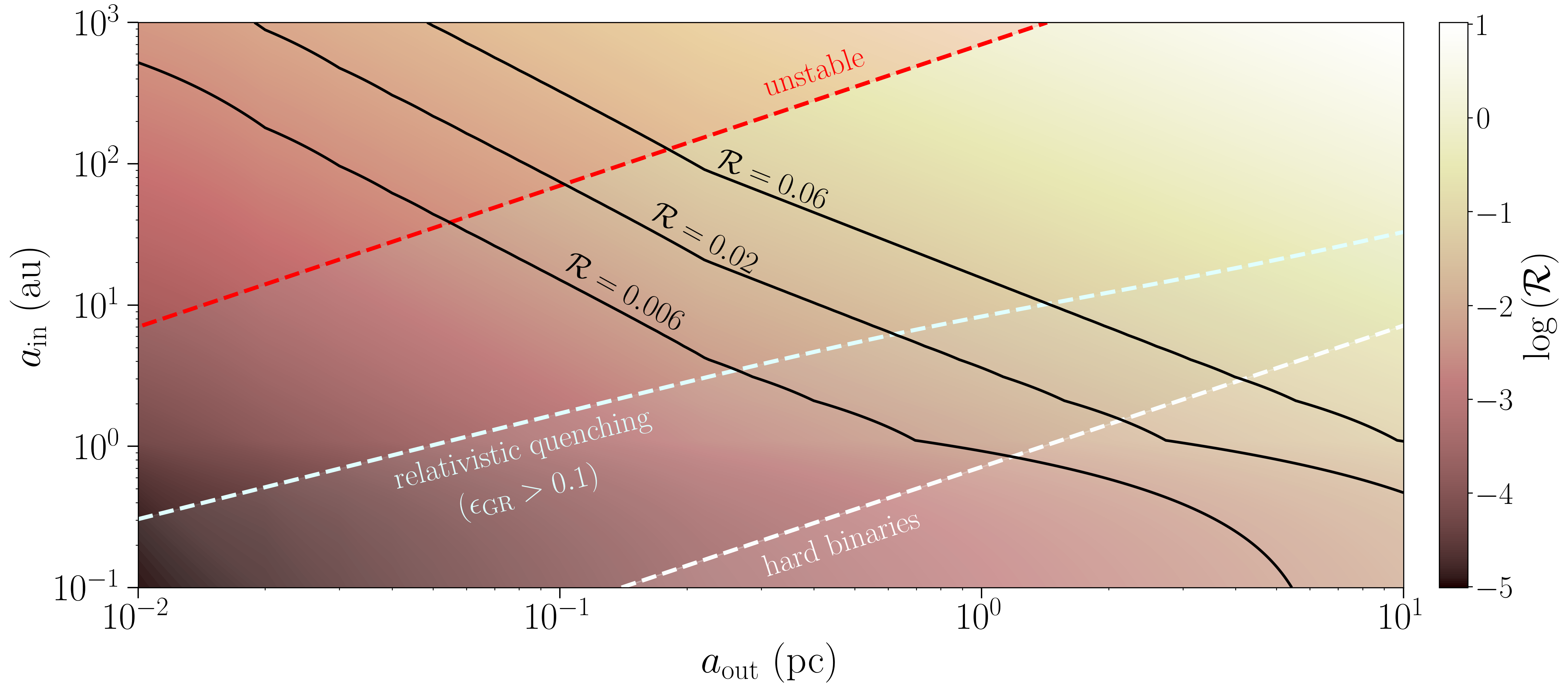}
\caption{Contours of the diffusion parameter $\mathcal{R}$ in Equation (\ref{eq:R}) as a function of $a\iB$ and $a\oB$, considering the fiducial scenario of $M\BH=4\times10^6M_\odot$, $M_p=1M_\odot$ and $M\bin=10M_\odot$. The boundary for unstable binaries is plotted as according to Equation (\ref{eq: stability}), while the limit for hard binaries is defined by $v_\text{kep}>\sigma$. We also plot the region in which relativistic quenching ($\epsilon_\text{GR}>0.1$) does not allow for mergers.}
\label{fig:R_contour}
\end{figure*}

In this section we analyze in more depth the diffusive effects of flybys on the binary evolution. 

As discussed in the previous section, the flybys impart random changes in all orbital elements ($e-i-\omega$), but it remains unclear whether the diffusion in $e_{\rm max}$ is dominated by the behavior of a single element or it is a collective process. As a controlled experiment, we show in Figure \ref{fig:pop_diffusion} the CDFs of the maximum eccentricities achieved by populations in which flyby encounters were replaced by random changes in individual orbital elements using the statistical distribution from N-body experiments. We clearly observe that diffusion in the inclination is the main effect driving binaries to high eccentricities. This translates into diffusion of $j_z$ as the dominant driver of maximum eccentricity grows. In turn, the diffusion of $\omega\iB$ and $e\iB$ adds little compared to the case where no diffusion occurs (i.e., tidal fields only).

Guided by these numerical experiments, we define a parameter $\mathcal{D}$ that quantifies the average diffusion rate of the specific angular momentum vector due to a single flyby encounter as 
\begin{equation}
    \mathcal{D}=\frac{\langle \|\Delta \vec{j}\|^2\rangle_{(f,b,i,\omega,\Omega)}}{t_\text{enc}}.
\end{equation}
This coefficient not only includes changes in the specific angular momentum $j=\sqrt{1-e^2}$, but also changes in its orientation which is relevant to model the changes in $j_z$. As explained in Appendix \ref{ap:R}, the $\langle \cdot \rangle$ average is carried over an ensemble of binaries with random orientations and phases relative to the stellar perturber. 

From this diffusion coefficient we define a dimensionless parameter $\mathcal{R}$ that quantifies the expected diffusion in $\vec{j}$ due to flybys after a secular timescale $\tau_0$ has occurred as
\begin{equation}
    \mathcal{R}=\sqrt{\mathcal{D}\tau_0}.
\end{equation}
Thus, the limits $\mathcal{R}\rightarrow0$ and $\mathcal{R}\rightarrow\infty$ refer to scenarios  completely dominated by tidal fields and by flybys, respectively. In Appendix \ref{ap:R} we show the full derivation of $\mathcal{D}$, in which we have used the impulse approximation to find an analytical expression for $\Delta\vec{j}$. From this we obtain
\begin{equation}
\label{eq:R}
\begin{split}
    \mathcal{R}\simeq&0.02\l(\frac{M\bin}{10M_\odot}\r)^{-\frac{1}{4}}\l(\frac{M_p}{M_\odot}\r)\l(\frac{M\BH}{4\times10^6M_\odot}\r)^{-\frac{1}{2}}\\
    &\times\l(\frac{v_p}{200\text{km/s}}\r)^{-\frac{1}{2}}\l(\frac{a\iB}{25\text{au}}\r)^{\frac{3}{4}}\l(\frac{a\oB}{0.3\text{pc}}\r)^{\frac{3}{2}}\\
    &\times\l(\frac{n_\star}{10^6\text{pc}^{-3}}\r)^{\frac{1}{2}}\l(\frac{f(\beta)}{1.8}\r)^{\frac{1}{2}},
\end{split}
\end{equation}
where $\beta = s/a\oB$ and $f(\beta)$ is defined in equation (\ref{eq:f(beta)}). The accuracy of this expression was verified numerically, by simulating the diffusion in a binary's angular momentum vector due to only flyby interactions using REBOUND. Figure \ref{fig:R_convergence} shows that the numerical equivalent for $\mathcal{R}$ also converges to $0.02$ when considering the fiducial scenario of $M\bin=10M_\odot$, $M_p=M_\odot$, $M\BH=4\times10^6M_\odot$, $a\iB=25$au and $a\oB=0.3$pc.

% As we can see in this expression, $\mathcal{R}$ grows dramatically as we increase $a\oB$. When this parameter is too high, the evolution of a binary becomes extremely diffusive as flyby effects dominate over tidal fields. This makes it much more difficult to reach the extreme eccentricities required for a merger, as will be discussed later in this section.

In general, we expect different types of evolution when varying $\mathcal{R}$, as it should become much more diffusive  and incoherent as this parameter grows. Figure \ref{fig:R_cases} shows this behavior, where we have evolved three binaries with different values of $\mathcal{R}$ by varying $M\BH$ in each case. All three binaries have initial parameters of $a\iB=25$au, $a\oB=0.3$pc, $e\iB=0.1$ and $e\oB=0$, and were evolved for $50\tau_0$ where $\tau_0$ is the initial secular timescale for each case according to equation (\ref{eq:tau_sec}). It is important to note that $\mathcal{R}$ increases with $\tau_\text{sec}$, therefore in a more diffusive scenario we will have more flybys per secular cycle as the cycles grow longer. We can see in the rightmost column how coherent cycles due to tidal fields become much more difficult to identify when $\mathcal{R}=0.06$, as opposed to the case where flybys start to become less important as shown in the first column where $\mathcal{R}=0.006$. Panel (d) more specifically shows a much less chaotic evolution in which we can easily distinguish coherent cycles with some perturbations such as extreme eccentricities and variations in the secular timescale, as opposed to panel (f) where the cycles are greatly perturbed by the flybys and hence such large eccentricities are not achieved. 

The bottom row shows the $z-$component of the angular momentum $j_z$ which should be a conserved quantity in a flyby-free scenario, however we can see how its value drifts due to the diffusion caused by flybys, especially as $\mathcal{R}$ increases. Due to the increasingly chaotic nature of the evolution of binaries in more diffusive regimes, it becomes much harder to reach the extreme eccentricities required to produce a merger, as can be seen in the case for $\mathcal{R}=0.06$. The large amount of flybys per secular timescale constantly resets the orbital parameters of the binary, preventing it from completing regular eccentricity cycles and hence decreasing the chances of achieving eccentricities of $1-e\lesssim10^{-4}$.

%As equation (\ref{eq:R}) shows, $\mathcal{R}$ also grows as we move further away from the center of the galaxy. This implies that environments such as the Oort cloud as studied by \cite{Heisler1986} should have a higher value for the diffusion parameter $\mathcal{R}$ in which binary evolution is flyby dominated, resulting in a more chaotic behaviour such as the one shown in the rightmost column of figure \ref{fig:R_cases}. In this scenario we will obviously have a much lower stellar density $n_\star$ which should, in principle, decrease the diffusion due to flybys. However, the dependency of $\mathcal{R}$ on $a\iB$ and $a\oB$ (both of which increase greatly as compared to the GC scenario) is more important than the dependency on $n_\star$, therefore we get an overall increase in diffusion and hence the production of mergers becomes more difficult.

\subsection{Experiments changing $\mathcal{R}$}
In Figure \ref{fig:merger_frac} we show the merger fractions of different binary populations as a function of $\mathcal{R}$, which was varied by changing the value of $M\BH$. These populations were all integrated for $\sim15$ Myr, which is roughly $50\tau_\text{sec}$ for a binary starting with $a\iB=25$au at $0.3$pc considering the fiducial case of $\mathcal{R}=0.02$. We consider any binary that reaches $1-e_\text{max}\leq10^{-4}$ to be a merger, as was discussed in section \S\ref{subsec:merger}. As mentioned previously, a less diffusive regime comes with shorter secular timescales; therefore, by integrating a population with $\mathcal{R}=0.006$ for $15$Myr we are actually letting them evolve for $\sim500\tau_\text{sec}$. However, as we keep lowering $\mathcal{R}$ by increasing $M\BH$, binaries start to become unstable very fast according to equation (\ref{eq: stability}) and hence the likelihood of them merging before they evaporate decreases. We also move into a regime in which the binaries only experience a few flyby interactions in each cycle, which could also contribute to decreasing the probability of reaching extreme eccentricity values as occurs in figure \ref{fig:merger_frac}. Our experiments suggest that there is an optimal value for $\mathcal{R}\sim 0.02$ at which the merger fraction peaks. 

\subsection{Understanding the optimal value of $\mathcal{R}$ to increase the merger fractions}

We provide an analytical explanation at the order-of-magnitude level of why there is a particular value for $\mathcal{R}$ at which the merger fraction peaks as shown in figure \ref{fig:merger_frac}.

We can write our merger condition as $j_\text{min}=\sqrt{1-e_\text{max}^2}$ reaching values below a critical specific angular momentum $j_\text{crit}$. The latter is set to $j_\text{crit}\simeq0.02$ in our simulations ($1-e_{\rm crit}=10^{-4}$). In the absence of flybys ($\mathcal{R}=0$), the fraction of mergers for initially circular and isotropic orbits is simply\footnote{The merger fraction is given by the probability that $j_\text{min}=\sqrt{5/3}|\cos I|<j_{\rm crit}$, which becomes $\sqrt{3/5}j_{\rm crit}$ for an isotropic distribution.} $\sqrt{3/5}j_{\rm crit}\simeq0.015 $ in our simulations. This changes slightly to $\simeq 0.52 j_{\rm crit}\simeq 0.01$ for an initial thermal distribution.

 As Figure \ref{fig:pop_diffusion} shows, the dominant effect of flybys on the merger rates comes from changes in inclinations, which corresponds to changes in $j_z$ at fixed $j$. Binaries that reach $|j_z|<j_\text{crit}$ can merge, provided that the binary lies in that range for at least a significant fraction of a secular timescale so it can be assisted by the tidal field. In a completely diffusive scenario, $j_z$ undergoes a random walk due to stellar encounters such that $\langle j_z^2\rangle=\mathcal{D}t$, with $\mathcal{D}$ a constant diffusive parameter. 
Thus, in a relaxation timescale $1/\mathcal{D}$ binaries outside the range $|j_z|<j_\text{crit}$ will be kicked in, monotonically increasing the fraction of mergers with $\mathcal{R}\propto \mathcal{D}$ consistent with the increase in Figure \ref{fig:merger_frac} below $\mathcal{R}\sim 0.02$.

As we continue to increase the diffusion rate due to stellar encounters $\mathcal{D}$, the merger fraction is expected to turn over when the binaries are kicked in and out of the $|j_z|<j_\text{crit}$ range in timescale shorter than the secular time. The 
critical diffusion timescale $t_\text{crit}$ to linger in this region is given by
 %We can then define $t_\text{relax}=1/D$ as the relaxation time for the angular momentum such that a binary will be kicked into or out of the $|j_z|<j_\text{crit}$ in a timescale $t_\text{diff}$ where
\begin{equation}
    t_\text{crit}\sim j_\text{crit}^2/\mathcal{D}.
\end{equation}

%As $j_\text{crit}$ is very small, we can rewrite

%\begin{equation}
%t_\text{diff}\sim j_\text{crit}^2t_\text{relax}\ll t_\text{relax}
%\end{equation}

%As we increase the strength of frequency of the encounters, the merger fraction increases as well as $t_\text{diff}$ becomes shorter. However, once it drops below the critical value $\tau_{\sec}$, the diffusion process becomes too strong to allow binaries to stay in the $|j_z|<j_\text{crit}$ long enough to merge, making the merger rate drop. Therefore, we can say that the merger fraction must peak when

Thus, we expect flybys to most efficiently boost the merger rate when $t_\text{crit}\sim \tau_0 $, a condition that can be expressed in terms of our definition of $\mathcal{R}$ as 
\begin{equation}
\label{eq: j_crit}
    \mathcal{R}=(\tau_0 \mathcal{D})^{1/2}\sim j_\text{crit},
\end{equation}
which would be the value of $\mathcal{R}$ for which the merger fraction peaks. In our simulations, this estimate predicts a peak value at $\mathcal{R}\sim 0.02$, which perfectly matches the observed peak in Figure \ref{fig:merger_frac}.

However, it is important to note that our simulations are not in a fully diffusive regime and  we observe sudden large jumps in $a\iB$ (see Figure \ref{fig:R_cases}) due to penetrative encounters ($b<a\iB$). Therefore, Equation (\ref{eq: j_crit}) serves as a general approximation to where the merger fraction should peak, but can be slightly off due to the non-diffusive nature of these penetrative flybys. In order to illustrate the approximate nature of our estimates, we repeat the numerical experiments for different values of $j_{\rm crit}$ in Figure \ref{fig:j_crit}. We observe that even though there is a value for $\mathcal{R}$ at which the merger fraction peaks, and it decreases for lower values of $j_\text{crit}$, the peaks do not occur exactly at $\mathcal{R}\sim j_\text{crit}$. 

\section{Lifting the relativistic quenching from flybys}
\label{sec:GR quenching}
%Figure \ref{fig:R_contour} shows a contour plot for $\mathcal{R}$ as a function of $a\iB$ and $a\oB$, in which we have plotted the values that $\mathcal{R}$ takes in the examples shown in figure \ref{fig:R_cases}. It is important to note that in this plot, equation (\ref{eq:R}) has been adapted such that $v_p$, $n_\star$ and $f(\beta)$ are written in terms of $a\oB$.

%Binaries residing in the bottom right corner of the plot are considered hard binaries and hence will only tend to get harder due to the Heggie-Hills Law. These binaries are also affected by relativistic quenching, which doesn't allow for mergers. However, there are also soft binaries that are subject to relativistic quenching, but are capable of escaping this dynamical regime by flyby interactions. 

Considering the properties of the GC, it is likely that a considerable amount of binaries residing in it will be subject to relativistic quenching. This presents an obstacle when producing mergers, as it does not allow binaries to reach extreme eccentricities. We can quantify the relative importance of relativistic precession using the following dimensionless parameter $\epsilon\GR$ as \citep{Petrovich2017}:
\begin{equation}
    \begin{split}
    \epsilon\GR&\equiv\f{\tau_\text{sec}}{\tau\GR}\\
        &=\f{4GM\bin^2a\oB^3(1-e\oB^2)^{3/2}}{c^2a\iB^4M\BH}\\
        &\times\l[1+\frac{M\cl}{M\BH}\f{2a\oB^{3-\gamma}}{(a\oB+s)^{3-\gamma}}\l(2-\f{3a\oB+s\gamma}{2(a\oB+s)}\r)\r]^{-1},
    \end{split}
\end{equation}
where binaries with values of $\epsilon\GR>0.1$ are considered as quenched.

When this parameter becomes smaller relativistic quenching can be lifted, allowing for a potential merger driven by the tidal fields. This can be achieved through the softening of binaries, as this will increase their secular timescale $\tau_{\rm sec}$. Conveniently, this paper has focused on the effect of flybys on binaries in the GC, which tend to soften said binaries and hence can effectively lift the effects of relativistic quenching. 

In figure \ref{fig:gr_case} we can see a specific example where softening driven by flybys lifts relativistic quenching and allows for a merger. The initial conditions of the simulated binary are $a\iB=1$au and $a\oB=0.3$pc, which makes it a soft binary severely quenched by GR with an initial dimensionless parameter $\epsilon\GR\sim 200$ . However, due to flyby interactions increasing $a\iB$, $\epsilon_\text{GR}$ reaches values smaller than 0.1 hence lifting relativistic quenching. It is clear that while subject to GR quenching the eccentricity of the binary's orbit remains practically constant at a low value, such that $a\iB\sim q\iB$. However once it escapes the GR quenching regime, extreme eccentricity oscillations are induced such that the pericentric distance reaches values as low as $10^{-3}$au, allowing for a potential merger.

%Figure \ref{fig:R_contour} shows the dynamical region where quenching forbids mergers, which is restricted to relatively tight binaries. However we know that flybys will tend to soften these binaries, pushing them upwards in the diagram, hence diminishing the effect of relativistic quenching and increasing the chances of them merging. This could be one of the sources of the newly found synergy between tidal fields and flybys; with tidal field evolution alone, soft binaries subject to relativistic quenching will never merge, therefore they are not merger candidates. But if they become softer due to flyby interactions, by lifting relativistic quenching effects they can potentially merge and thus increase the overall pool of merger candidates in a population. 

\section{Discussion}
\label{sec:discussion}

We have shown that stellar flybys can greatly enhance the merger rates of binaries in the centers of galaxies by assisting the eccentricity growth to extreme values that are driven by the tidal torques exerted by the SMBH and the central cluster. 

%In this work we have studied the combined effect of cluster tidal fields and flyby interactions on the evolution of binaries in the galactic center. Our main result is the clear synergy between the two physical processes at driving binaries into extremely high eccentricities ($1-e_\text{max}\lesssim10^{-4}$ and their subsequent merger.

%This synergy can enhance the rate of stellar mergers by over an order of magnitude compared to the cases where flybys are ignored.

More specifically, our results demonstrate that the diffusive effects of flybys and the torquing of binary's orbits due to tidal fields produce comparable merger rates (within a factor of 3; see Figure \ref{fig:cdf-all}). In turn, for regimes in which the diffusion rate is much slower than the tidal field torquing rate ($\mathcal{R}\sim 0.01$; see \S\ref{sec:tides vs flybys}) flybys added to the tidal fields are able to produce a merger rate $\sim30$ times larger that the one produced by tidal fields alone. This is due to the chaotic behavior introduced by flyby interactions, as they constantly reset the binary's orbital parameters and consequently can produce random and dramatic changes in the orbit. Most importantly, the diffusion of the $z-$component of the angular momentum $j_z$.

We also find that the softening of GR quenched binaries due to flyby interactions allows them to escape the relativistic quenching regime, thus expanding the available phase-space for mergers.

%The consideration of flyby interactions plays a significant role in binary evolution, specially when leading to mergers. These interactions have the particular effect of resetting the binary's inner orbital parameters constantly, disrupting the smooth secular evolution produced by tidal fields. This adds a considerable chaos component to the overall evolution of binaries, which contributes in helping them achieve extreme eccentricites when added to tidal field effects.

%We find that this synergy increases merger rates in 100 secular timescales from $1\%$ ($3\%$) when considering only tidal field (flyby) effects, to $\sim37\%$ when taking into account both evolution channels simultaneously. We also demonstrate that this is a cumulative process, as more binaries are able to achieve extreme eccentricities if we let them evolve for a longer period of time. 

\subsection{Dynamical regimes in the galactic center}
\label{subsec:dynamical regimes GC}

In Figure \ref{fig:R_contour} we show the different dynamical regimes as a function of $a\iB$ and $a\oB$, along with a contour plot of $\mathcal{R}$. It is important to note that in this plot, Equation (\ref{eq:R}) has been adapted such that $v_p$, $n_\star$ and $f(\beta)$ are written in terms of $a\oB$. We can see that for binaries that are extremely tight ($a\iB\lesssim0.1$au), $\mathcal{R}$ reaches very low values which implies an almost insignificant diffusion in the angular momentum. If these binaries are subject to relativistic quenching, even if they are soft, it is unlikely that flyby interactions will have enough impact on their orbit to push them out of this regime. Considering this, the lifting of GR quenching due to flyby interactions is probably most efficient when $a\iB\gtrsim0.1$ au

%We derived a dimensionless parameter $\mathcal{R}$ to quantify the diffusion of angular momentum in one secular timescale due to flyby interactions, finding an optimal value that enhances mergers at $\mathcal{R}\approx0.006$. This scenario can be seen in the galactic center, with the exception of binaries that are too close to the SMBH. In their case tidal fields will dominate over flybys, added to the fact that mergers become more unlikely due to GR quenching. However, this region isn't extremely relevant as it is confined to either very small values of either $a\oB$ or $a\iB$, therefore in general our results further reinforces that the centers of galaxies can produce compact object binary mergers very efficiently.

Our results suggest that previous studies such as \cite{Petrovich2017} and \cite{Hoang2018} (who considered a regime similar to our own with $a\oB\lesssim0.5$pc and $\epsilon\GR\lesssim0.1$) have underestimated the effect of interactions with field stars, presenting it as merely a limiting evaporation timescale that should inhibit merger rates. Our findings on the other hand, suggest that flyby interactions are actually quite efficient when producing mergers, much more so than tidal fields alone. This is due mainly to the diffusion in the angular momentum vector, quantified by our dimensionless parameter $\mathcal{R}$, that for these previous works would be of order $\mathcal{R}\sim0.01$. This, combined with the ability to lift relativistic quenching and hence increase the phase-space for mergers (which was not taken into account in the mentioned studies either), provides strong evidence for the importance of flybys in producing mergers in the GC and shows that their effects must be accounted for.

\subsection{Other astrophysical environments}
Our results are applicable the evolution of wide binaries in various astrophysical environments. Next we discuss our results in the context of previous results.

\paragraph{Solar neighborhood} The dynamical evolution of wide binaries is equivalent to that of comets in the Oort Cloud studied by \cite{Heisler1986}. This is a different regime than the one considered in this paper, where the evolution should be flyby dominated (higher value for $\mathcal{R}$) due to the orbits of binaries being wider and having longer secular timescales. We can write an expression for $\mathcal{R}$ in this environment as follows

%This makes it difficult to see any kind of synergy between tidal fields and flybys, as opposed to the case of the galactic center. The fact that the secular timescales in this environment are much longer also makes it hard for this synergy to present itself, considering that the binaries can only undergo a few cycles in one Hubble time. This means that the eccentricity peaks are only achieved a few times, making it much less likely to reach extreme eccentricity values.

\begin{equation}
\label{eq:R_oort}
\begin{split}
    \mathcal{R}\simeq&9\l(\frac{M\bin}{M_\odot}\r)^{-\frac{1}{4}}\l(\frac{M_p}{0.162M_\odot}\r)\l(\frac{v_p}{40\text{km/s}}\r)^{-\frac{1}{2}}\\
    &\times\l(\frac{a\iB}{2.5\times10^4\text{au}}\r)^{\frac{3}{4}}\l(\frac{n_\star}{1.14\text{pc}^{-3}}\r)^{\frac{1}{2}}
    \l(\frac{\rho_0}{0.185M_\odot\text{pc}^{-3}}\r)^{-\frac{1}{2}},
\end{split}
\end{equation}
where we have considered the secular timescale as

\begin{equation}
\label{eq:tau_g}
    \tau_g = \frac{M\bin^{1/2}}{2\pi\rho_0G^{1/2}a\iB^{3/2}},
\end{equation}
with $\rho_0=0.185\pm0.02M_\odot\text{pc}^{-3}$ being the local density as used by \cite{Heisler1986}.

We can see that for an average comet in the Oort Cloud we get $\mathcal{R}\simeq 9$ as opposed to the $\mathcal{R}\simeq 0.02$ obtained for compact object binaries in the GC. This implies that in the Oort Cloud flybys dominate over tidal fields, as the diffusion rate from these interactions is $\sim9$ times faster than the torquing due to the tidal field. This is mainly due to the difference in the amount of flybys in a secular timescale, $\tau_0/t_\mathrm{enc}$. As shown in equation (\ref{eq:n_fbs}), we get about 68 encounters for an average binary in the GC, while in the solar neighbourhood the average comet suffers $\sim2\times10^4$ flyby encounters (considering equation (\ref{eq:tau_g}) as $\tau_0$ and $t_\mathrm{enc}\simeq0.015\mathrm{Myr}$ as used by \citealt{Heisler1986}). This increase in the amount of close encounters inevitably results in more diffusion of the angular momentum $\vec{j}$, hence the higher value of $\mathcal{R}$ obtained in this regime. Consistently, the numerical experiments by \citet{Heisler1986} show the secular evolution is largely detuned due to the stellar flybys, making it hard to complete a clean secular cycle in the $\omega-e$ space (see figure 4 therein). This in turn makes it difficult to achieve the high eccentricities required by a merger, as is discussed by \cite{stegmann2024}.

\paragraph{Globular clusters} A recent study by \cite{Rasskazov2023} looked at the evolution of compact object binaries when subject to tidal fields and stellar encounters in globular clusters, similar to the work done throughout this paper. The main difference is that these globular clusters are less dense and do not contain a central SMBH as the NSCs we have considered in this study, which means that they have a much lower velocity dispersion. This results in a higher $a\iB$ hard-soft limit for binaries, implying that many of the binaries in these environments are actually hard, unlike our work. This is particularly important when speaking of GR quenching, as we have found that flybys are able to diminish this effect only when they are capable of softening binaries. In the case of hard binaries, flybys should tend to promote GR quenching rather than lift it, further inhibiting merger rates.

%the authors find that for wide enough binaries (e.g., >1000 au) flybys have the ability of widening the orbit such that GR precession is no longer an obstacle. In turn, tight enough binaries (e.g., <100 au) can harden in time and quench the eccentricity effect driven by the tidal fields.
By looking at the behavior in the various evolution examples in \cite{Rasskazov2023}, we suggest that similar to the galactic field, the diffusion parameter $\mathcal{R}$ is large (dominated by flybys). We note that equation (\ref{eq:R}) is not applicable here because the encounters are generally far from being impulsive given the low dispersion velocities. More work would need to be done to extend the analysis to these environment, including flybys that can be secular in nature (e.g., \citealt{heggie_rasio1996,hamers2019}).

%However, as the clusters they considered do not contain a central SMBH, the secular timescales for binary evolution will be much longer, resulting in a large amount of flyby encounters and hence much more diffusion in the angular momentum vector, independent of whether the encounters are impulsive or not.

%i) Because of the lower dispersion velocities in GCs compared to galactic centers, the encounters are less impulsive.
%This is unlike the galactic center wher the 

\section{Conclusions}
\label{sec:conclusions}
%In this work, we have shown the relevance of flyby interactions in GC binary merger rates, finding for the first time a dramatic synergy with tidal fields that can increase the 
In this work we have studied the combined effect of cluster tidal fields and flyby interactions on the evolution of binaries in the galactic center. Our main result is the dramatic synergy between these two physical processes at driving binaries into extremely high eccentricities ($1-e_\text{max}\lesssim10^{-4}$) and their subsequent merger.

Although stellar flybys tend to increase the semi-major axis of the binaries thus inhibiting the mergers, they also give rise an stochastic evolution of other orbital elements that can dramatically shrink the pericenter distance before the binary becomes unbound. This dynamics has been previously studied showing that it could lead to mergers of wide binaries in the galactic field, consistent with our results for the galactic center including only flybys that drive mergers in few a percent of the systems. We extend these results and show that this rate increases by an order of magnitude when flybys act in concert with the static tidal field from the SMBH and the central cluster. 

%We knew already that in this environment, binaries are subject to strong tidal fields that efficiently torque its orbit, producing extremely high eccentricities that lead to mergers. Close interactions with field stars, on the other hand, tend to soften the orbit of these binaries, potentially evaporating them and preventing mergers from occurring.  However, in this work we found that although flybys can indeed cause evaporation, they have the particular effect of resetting the binary's inner orbital parameters constantly, disrupting the smooth secular evolution produced by tidal fields. This adds a considerable chaos component to the overall evolution of binaries, which contributes in helping them achieve extreme eccentricites when added to tidal field effects. In other words, the overall effect of flyby interactions on binaries is the enhancement of merger rates. When combining tidal field and flyby effects, merger rates increase by a factor of $\sim30$ $(10)$ as compared to models that consider tidal fields (flybys) on their own. 

We find that the main cause of this synergy is the persistent tidal field-driven eccentricity excitation that is enhanced by the diffusion of $j_z$ due to flyby encounters. We calculated the diffusion rate of the angular momentum vector $\vec{j}$  due to flybys using the impulse approximation and assess the various regimes with a dimensionless parameter $\mathcal{R}$ (Eq. \ref{eq:R}) that measures the expected diffusion within a secular timesale (i.e, torquing timescale due to tidal fields). Our experiments suggest that the merger rates peak in the range $\mathcal{R}\sim 0.01-0.1$, meaning when the diffusion rate is $\sim10-100$ times more slow than the torquing due to the tidal field. This regime is typical of the central parsec of our GC (see Figure \ref{fig:R_contour}), while other environments such the solar neighborhood and stellar clusters tend to have much higher values of $\mathcal{R}$.

We also show that the gradual softening of binaries in the GC can lift the relativistic quenching of initially tight binaries, thus expanding the available phase-space for mergers. This is also likely to contribute to the previously discussed synergy, given that by increasing the amount of binaries that can potentially merge in a population we can expect a higher merger rate.

In summary, we conclude that despite the gradual softening of binaries due to stellar encounters, these greatly enhance merger rates in GCs by promoting the tidal field driven eccentricity excitation This behavior has been ignored in previous works and further reinforces that galactic centers are ideal environments for the production of compact object binary mergers.\\

\noindent We would like to thank Antranik Sefilian, Carolina Charalambous, Diego Mu\~noz, Mark Dodici, and Scott Tremaine for useful discussions. We are also grateful to an anonymous referee for providing helpful comments on the maniscript.
CP acknowledges support from CATA-Basal AFB-170002, ANID BASAL project FB210003, FONDECYT Regular grant 1210425, CASSACA grant CCJRF2105, and ANID+REC Convocatoria Nacional subvencion a la instalacion en la Academia convocatoria 2020 PAI77200076.
CH is supported by the John N. Bahcall Fellowship Fund at the Institute for Advanced Study.

\bibliography{refs}
\appendix

\section{Impulse approximation}
\label{ap:IA}

From \cite{CollinsSari2008}, we can write the variation in the velocity (impulse) $\Delta\Vec{v_i}$ of one component of the binary with mass $m_i$ ($i=\{1,2\}$) considering an impact parameter $\Vec{b_i}$ as follows:

\begin{equation}
\label{eq:deltav}
    \Delta\Vec{v_i} = \frac{2Gm_p}{v_pb_i}\hat{b_i}.
\end{equation}

The impact parameter and the velocity of the perturber must be orthogonal such that $\Vec{b}_i\cdot\Vec{v}_p=0$. The trajectory of the perturber is approximated to a straight line, such that $\Vec{r}(t) = \Vec{b}_i+\Vec{v}_pt$. The perturber's velocity is relative to only one component of the binary; therefore, we can relate both impact parameters $b_1$ and $b_2$ through the equation $\Vec{b_2} = \Vec{b}_1-\Vec{r}+\hat{v}_p(\Vec{r}\cdot\hat{v}_p)$ considering that $\Vec{b_2}$ must also be perpendicular to $\Vec{v}_p$.  
The overall change in relative velocity can then be expressed as $\Delta\Vec{v}=\Delta\Vec{v}_1-\Delta\Vec{v}_2$, where both $\Delta\Vec{v}_1$ and $\Delta\Vec{v}_2$ obey the equation \ref{eq:deltav}.  

%We can consider the specific orbital energy for a 2 body problem defined as:
%\begin{equation}
%    \varepsilon=\frac{\Vec{v}^2}{2}-\frac{GM_{bin}}{r}=-\frac{GM_{bin}}{2a_{in}}
%\end{equation}
%Due to the impulse approximation, we consider that the distance between both components of the binary $\Vec{r}$ doesn't vary during the encounter. Therefore, we can calculate the change in the specific orbital energy as:
%\begin{equation}
%    \Delta\varepsilon=\frac{(\Vec{v}+\Delta\Vec{v})^2}{2}-\frac{v^2}{2} = \frac{\Delta v^2+2\Vec{v}\cdot\Delta\Vec{v}}{2}
%\end{equation}
We can define $\Delta\Vec{v}$ as the sum of the velocity components perpendicular and parallel to the instantaneous orbital velocity; $\Delta\Vec{v} = \Delta v_{\perp}+\Delta v_{\parallel}$. We can rewrite it using equation \ref{eq:deltav} as
%\begin{equation}
%\label{eq:deltaa}
%    \begin{split}
%        \Delta a &= -a\frac{\Delta\varepsilon}{\varepsilon}\\
%        &= \frac{a^2}{GM_{bin}}(\Delta v^2+2\Vec{v}\cdot\Delta\Vec{v})
%    \end{split}
%\end{equation}

\begin{equation}
    \Delta\Vec{v} = \frac{2GM_p}{v_p}\left(\frac{\hat{b}_1}{b_1}-\frac{\hat{b}_2}{b_2}\right)
\end{equation}

\renewcommand{\thefigure}{A\arabic{figure}}
\setcounter{figure}{0}

\begin{figure}
    \centering
    \includegraphics[width=\linewidth]{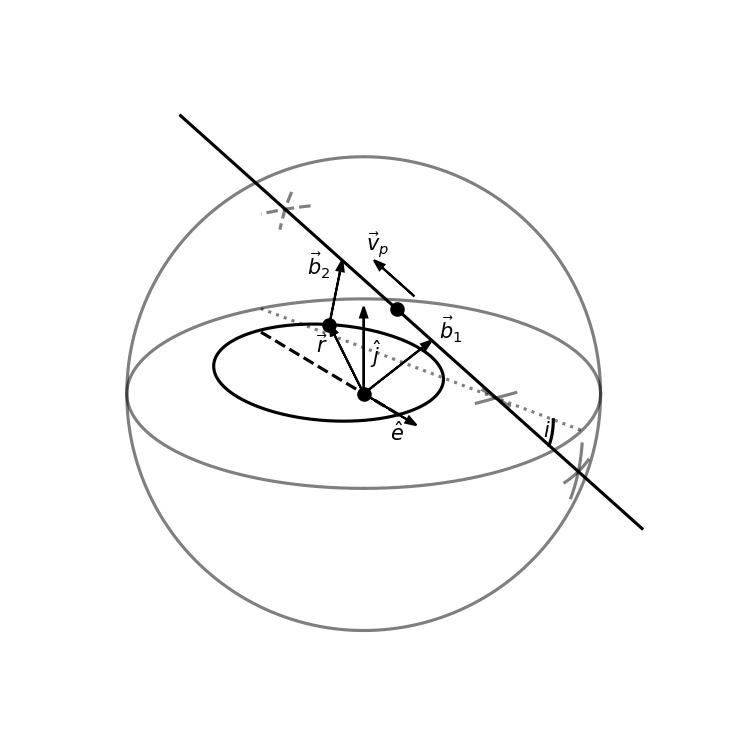}
    \caption{Geometry used to calculate $\hat{r}\times\hat{b}_1$ and $\hat{r}\times\hat{b}_2$. The orbit lies on the equatorial plane of a sphere for reference. The flyby trajectory is approximated to a straight line, and the gray crosses indicate the points where it intersects either the sphere or its equator.%, the dashed cross indicates such point is beyond the plane of the paper.
    }
    \label{fig:geometry}
\end{figure}

When the perturbing potential is axisymmetric (as is the case in this scenario), the inner angular momentum and eccentricity vectors $\Vec{j}$ and $\Vec{e}$ must always be orthogonal such that $\Vec{j}\cdot\Vec{e}=0$. Therefore, if there is a variation in the eccentricity vector due to fly-by encounters, there must also be a variation in $\Vec{j}$. We can write the change in the angular momentum vector as
\begin{equation}
    \Delta\Vec{j}=\frac{\Vec{r}\times\Delta\Vec{v}}{\sqrt{GM\bin a\iB}}
\end{equation}
which, using \ref{eq:deltav}, results in 

\begin{equation}
    \Delta\Vec{j}=\sqrt{\frac{G}{M\bin a\iB}}\frac{2M_p}{v_p}\left(\frac{\Vec{r}\times\hat{b}_1}{b_1}-\frac{\Vec{r}\times\hat{b}_2}{b_2}\right) \label{eq:Dvecj}.
\end{equation}

\subsection{Diffusion coefficient analysis}
\label{ap:R}

It would be of interest to analyse the relative contributions of the tidal and the impulse elements on the evolution of the angular momentum of the inner binary, for it is an indicator of how likely the binary is to merge at a given moment. To this purpose, we define a diffusion coefficient $\mathcal{D}\equiv\langle\|\Delta j\|^2\rangle_{(f,b,i,\omega,\Omega)}/t_\text{enc}$. From equation \ref{eq:Dvecj}, we can write $\|\Delta\vec{j}\|$ as

\begin{equation}
    \|\Delta\vec{j}\|= \sqrt{\frac{G}{M\bin a\iB}}\frac{2M_p}{v_p}r\left\|\frac{\hat{r}\times\hat{b}_1}{b_1}-\frac{\hat{r}\times\hat{b}_2}{b_2}\right\|,
\end{equation}
where the term involving the cross products can be evaluated numerically by averaging over $10^6$ different values calculated using random impact parameters $b_i$ and  orbital parameters $f$, $\omega$, $i$ and $\Omega$ so as to cover all possible stages of the binary's orbit. Using this numerical average, we can write $\langle\|\Delta j\|^2\rangle_{(f,b,i,\omega,\Omega)}$ as

\begin{equation}
\begin{split}
\label{eq:delta_j2}
    \langle\|\Delta \vec{j}\|^2\rangle _{(f,b,i,\omega,\Omega)}
    \approx&6.33\times10^{-6}\l(\frac{M\bin}{10M_\odot}\r)^{-1}\l(\frac{M_p}{M_\odot}\r)^{2}\\
    &\l(\frac{v_p}{200\text{km/s}}\r)^{-2}\l(\frac{a\iB}{25\text{au}}\r),
\end{split}
\end{equation}
where we have used the average $\langle r^2\rangle=a\iB^2(1+\frac{3}{2}e^2)=\frac{7}{4}a\iB^2$, considering that $\langle e^2\rangle=0.5$ due to the initial thermal distribution assumed.

\subsection{Number of flybys}
\label{ap:N}

The number of interactions in a secular timescale can be calculated as $\tau_0/t_\text{enc}$, following equations (\ref{eq:tau_sec1}) and (\ref{eq:enc_time})
\begin{equation}
\begin{split}
\label{eq:n_fbs}
    \frac{\tau_0}{t_\text{enc}}&\approx67.5 \l(\frac{M\bin}{10M_\odot}\r)^{\frac{1}{2}}\l(\frac{M\BH}{4\times10^6M_\odot}\r)^{-1}\\
    &\times\l(\frac{a\oB}{0.3\text{pc}}\r)^3\l(\frac{a\iB}{25\text{au}}\r)^{\frac{1}{2}}\l(\frac{n_\star}{10^6\text{pc}^{-3}}\r)\\
    &\times
    \l(\frac{v_p}{200\text{km/s}}\r)
    \l(\frac{f(\beta)}{1.8}\r),
\end{split}
\end{equation}
where $\beta = s/a\oB$, and 

\begin{equation}
\label{eq:f(beta)}
    f(\beta) = \l[\frac{1}{2}+\frac{4(1+3\beta)}{(1+\beta)^3}\r]^{-1}.
\end{equation}

%Combining this with equation (\ref{eq:delta_j2}) results in %the final expression

%\begin{equation}
%\begin{split}
%    \mathcal{R}=&0.001\l(\frac{M\bin}{10M_\odot}\r)^{-\frac{1}{2}}\l(\frac{M_p}{M_\odot}\r)^{2}\l(\frac{M\BH}{4\times10^6M_\odot}\r)^{-1}\\
%    &\times\l(\frac{v_p}{200\text{km/s}}\r)^{-1}\l(\frac{a\iB}{25\text{au}}\r)^{-\frac{1}{2}}\l(\frac{a\oB}{0.3\text{pc}}\r)^{3}\\
%    &\times\l(\frac{n_\star}{10^6\text{pc}^{-3}}\r)\l(\frac{f(\beta)}{1.8}\r)
%\end{split}
%\end{equation}

\subsection{Numerical experiments for $\mathcal{R}$}
\begin{figure}
\centering
\includegraphics[width=\linewidth]{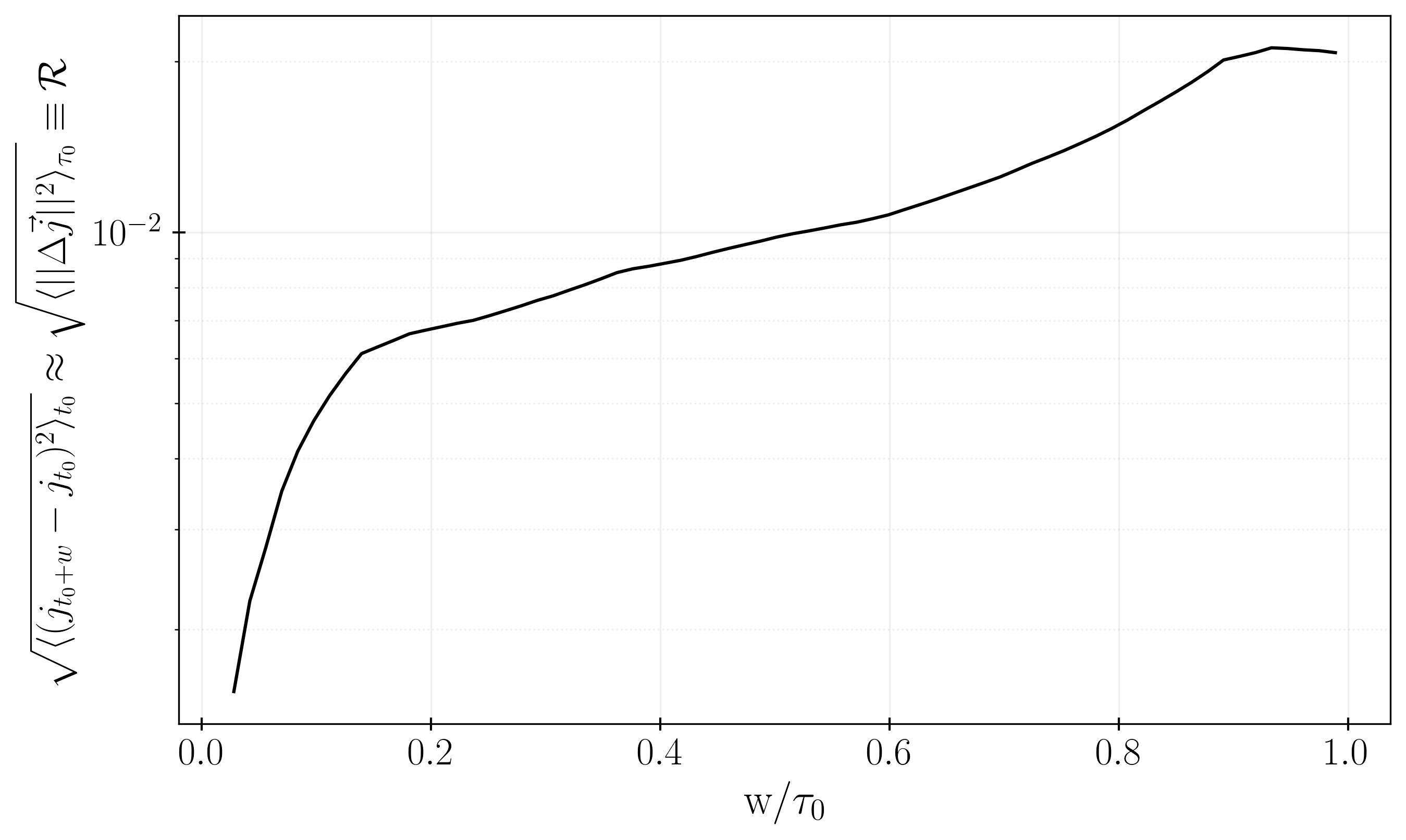}
\caption{Mean square displacement of $\sqrt{\langle\|\Delta\vec{j}\|^2\rangle}$ as a function of time window width $w/\tau_0$.}
\label{fig:R_convergence}
\end{figure}

In order to confirm that the expression derived for $\mathcal{R}$ using the impulse approximation is accurate, we simulated a binary that evolves only by flyby interactions using REBOUND. From this numerical experiment, in which we considered $M\bin=10M_\odot$, $M_p=M_\odot$, $M\BH=4\times10^6M_\odot$, $a\iB=25$au, $a\oB=0.3$pc and $v_p=200$kms$^{-1}$, we calculated the mean displacement of $\|\vec{j}\|$ in different time windows until finding a value at which it converges after one secular timescale $\tau_0$. This is shown in Figure \ref{fig:R_convergence}, where we can clearly see how $\mathcal{R}$ converges to $\approx0.02$ after one secular timescale, which matches the value we retrieve using equation (\ref{eq:R}) when considering the same initial conditions. In order to keep this simulation as similar as possible to the derivations using the impulse approximation, we do not consider the changes in $a\iB$ after each encounter, as this would change the amount of flybys per secular timescale.

\section{Merger fraction and $j_\text{crit}$}

To show that the optimal value of $\mathcal{R}$ at which the merger fraction peaks is proportional to the $j_\text{crit}$ chosen to classify mergers, a set of populations was simulated at different values for $\mathcal{R}$. Their merger fractions for three different values of $j_\text{crit}$ are plotted in figure \ref{fig:j_crit} as a function of $\mathcal{R}$, and we can see that for smaller $j_\text{crit}$ the peak value tends towards a smaller $\mathcal{R}$ as well.

\setcounter{figure}{0}
\makeatletter 
\renewcommand{\thefigure}{B\@arabic\c@figure}
\makeatother
\begin{figure}
\centering
\includegraphics[width=\linewidth]{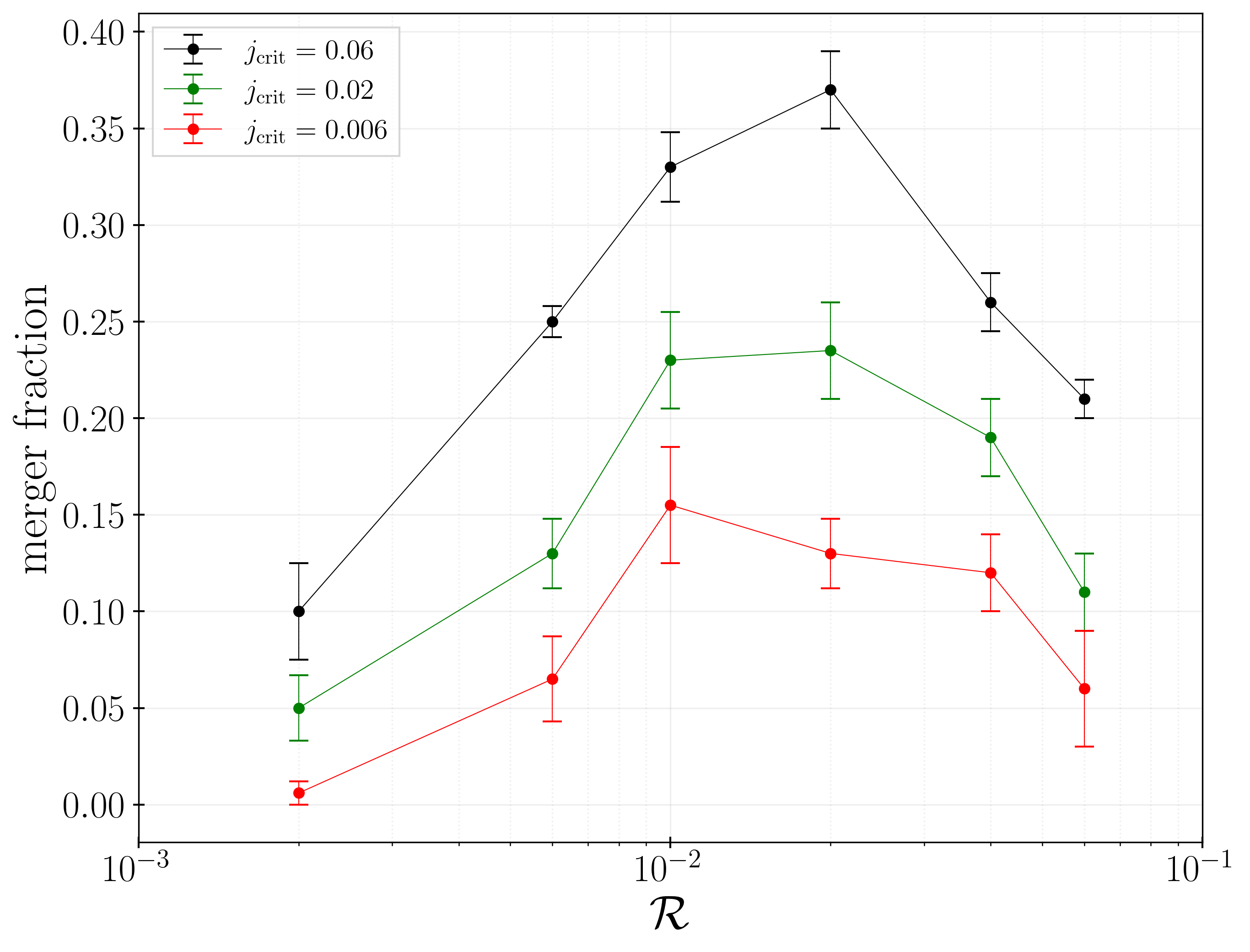}
\caption{Merger fraction as defined by different values of $j_\text{crit}$ as a function of $\mathcal{R}$.}
\label{fig:j_crit}
\end{figure}

\end{document}